\documentclass[sigconf,authorversion]{acmart}

\AtBeginDocument{%
	\providecommand\BibTeX{{%
			\normalfont B\kern-0.5em{\scshape i\kern-0.25em b}\kern-0.8em\TeX}}}

\acmConference[ACSAC '20]{Annual Computer Security Applications Conference}{December 7-11, 2020}{Austin, Texas}
\copyrightyear{2020}
\acmYear{2020}
\setcopyright{acmlicensed}\acmConference[ACSAC 2020]{Annual Computer Security Applications Conference}{December 7--11, 2020}{Austin, USA}
\acmBooktitle{Annual Computer Security Applications Conference (ACSAC 2020), December 7--11, 2020, Austin, USA}
\acmPrice{15.00}
\acmDOI{10.1145/3427228.3427295}
\acmISBN{978-1-4503-8858-0/20/12}

\acmBadgeR[https://www.acsac.org/2020/submissions/papers/artifacts/]{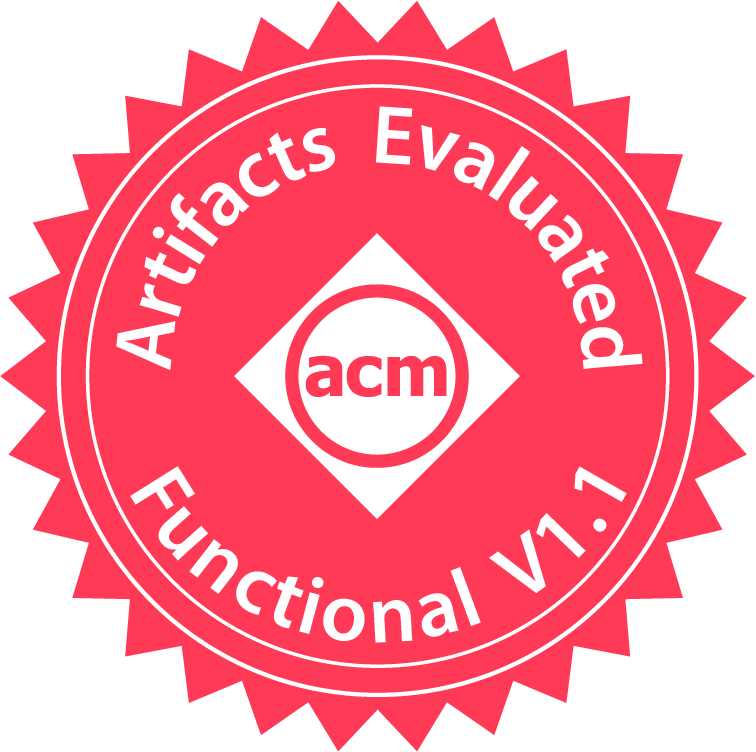}

\usepackage{acmart-taps}
\usepackage{float}
\usepackage{listings}
\definecolor{dkgreen}{rgb}{0,0.6,0}
\definecolor{gray}{rgb}{0.5,0.5,0.5}
\definecolor{lightgray}{rgb}{0.8,0.8,0.8}
\definecolor{mauve}{rgb}{0.58,0,0.82}
\graphicspath{{resources/}}

\begin{document}

\title{Dragonblood is Still Leaking: \\
Practical Cache-based Side-Channel in the Wild}

\author{Daniel De Almeida Braga}
\email{daniel.de-almeida-braga@irisa.fr}
\affiliation{%
	\institution{Univ Rennes, CNRS, IRISA}
	\city{Rennes}
	\country{France}
}

\author{Pierre-Alain Fouque}
\email{pierre-alain.fouque@irisa.fr}
\affiliation{%
	\institution{Univ Rennes, CNRS, IRISA}
	\city{Rennes}
	\country{France}
}

\author{Mohamed Sabt}
\email{mohamed.sabt@irisa.fr}
\affiliation{%
	\institution{Univ Rennes, CNRS, IRISA}
	\city{Rennes}
	\country{France}
}

\begin{abstract}
Recently, the Dragonblood attacks have attracted new interests on the security of WPA-3 implementation and in particular on the Dragonfly code deployed on many open-source libraries. One attack concerns the protection of users passwords during authentication. In the Password Authentication Key Exchange (PAKE) protocol called Dragonfly, the secret, namely the password, is mapped to an elliptic curve point. This operation is sensitive, as it involves the secret password, and therefore its resistance against side-channel attacks is of utmost importance. Following the initial disclosure of Dragonblood, we notice that this particular attack has been partially patched by only a few implementations. 

In this work, we show that the patches implemented after the disclosure of Dragonblood are insufficient. We took advantage of state-of-the-art techniques to extend the original attack, demonstrating that we are able to recover the password with only a third of the measurements needed in Dragonblood attack. We mainly apply our attack on two open-source projects: iwd (iNet Wireless Daemon) and FreeRADIUS, in order underline the practicability of our attack. Indeed, the iwd package, written by Intel, is already deployed in the Arch Linux distribution, which is well-known among security experts, and aims to offer an alternative to wpa\_supplicant. As for FreeRADIUS, it is widely deployed and well-maintained upstream open-source project. We publish a full Proof of Concept of our attack, and actively participated in the process of patching the vulnerable code. Here, in a backward compatibility perspective, we advise the use of a branch-free implementation as a mitigation technique, as what was used in hostapd, due to its quite simplicity and its negligible incurred overhead.
\end{abstract}




\maketitle

\section{Introduction} \label{sec:introduction}
%!TEX root = article.tex

\subsection{Context and Motivation}
Fourteen years after the implementation of WPA2, the WPA3 protocol was introduced by the Wi-Fi Alliance in early January 2018. WPA3 was much anticipated after severe weaknesses identified in WPA2 in Fall 2017 using key reinstallation attacks (KRACKs)~\cite{DBLP:conf/ccs/VanhoefP17}. WPA3 aims at improving authentication and encryption during connections. Indeed, it replaces Pre-Shared Key (PSK) authentication by WPA3-SAE (Simultaneous Authentication of Equals). Unlike PSK, SAE resists offline dictionary attacks; namely the only way for an attacker to guess a password is through repeated trials. A security requirement is that each trial must only reveal one single password, thereby forcing online attacks that can be easily mitigated through, for instance, limiting authentication attempts. Thus, SAE, which is a variant of the Dragonfly handshake, is considered as a major addition to WPA3. SAE is defined in the standard IEEE 802.11-2016~\cite{7786995}, that implements a slight variant the Dragonfly RFC defined in~\cite{rfc7664}.

% If we talk about doubt, there was a large polemic around the whole standardization (namely because of the password conversion). Maybe cite them
Nevertheless, some researchers cast some doubt on the guarantees promised by SAE and Dragonfly~\cite{IETF:mail_archive/ipsec/NEicYFDYJYcQuNdknY0etLyfITA,IETF:mail_archive/cfrg/_BZEwEBBWhOPXn0Zw-cd3eSV6pY,IETF:mail_archive/cfrg/LsFX5Qqw53dTUmSsUOooLca5FHg,IETF:mail_archive/tls/A_SfHI4BsdAi4miklBs3TvUbu-Y}. In 2019, Vanhoef and Ronen identified a set 
of vulnerabilities in WPA3 implementations, especially against its password-encoding method~\cite{DBLP:conf/sp/VanhoefR20}. 
Along with the vulnerability, they present a collection of attacks, along with appropriate mitigations. Among their attacks, some exploit 
both timing and cache side-channels in order to leak some information. Then, they show how the leak is related to the targeted password, and 
mount an offline dictionary attack accordingly. The disclosure of Dragonblood is unfortunate to the Wi-Fi Alliance that has just got its biggest 
update in 14 years. However, this did not discourage vendors to continue their WPA3 adoption, especially that KRACKs of WPA2 is more 
serious, since it concerns the standard itself, while Dragonblood mainly leverages implementation weaknesses related to side-channel leaks. 
In response, the Wi-Fi Alliance published some implementation guidance to be followed by manufacturers~\cite{wpa3_security_considerations} 
to ensure secure backward compatible WPA3's implementations. Authors in~\cite{DBLP:conf/sp/VanhoefR20} cast doubts on the 
endorsement of some backwards-compatible side-channel defenses, especially in the context of resource-constrained devices because of 
their high overhead. Moreover, they argue that a secure implementation of the countermeasures is an arduous task.

In this paper, we focus on the recommendations related to \textit{Cache-Based Elliptic Curve Side-Channels} in~\cite{wpa3_security_considerations}, which address mitigations to the set of Dragonblood vulnerabilities related to cache-based attacks. Two mitigations are underlined: (i) performing extra dummy iterations on random data, and (ii) blinding the calculation of the quadratic residue test. For the first mitigation, the RFC 7664~\cite{rfc7664} recommends that 40 iterations are always executed even if the password was successfully encoded requiring fewer iterations. Concerning the second mitigation, a blinding scheme is suggested for the function determining whether or not a value is a quadratic residue modulo a prime.

\subsection{Our Contribution}
In our paper, we show that such countermeasures are not enough to defend against cache-based side-channel attacks. In fact, these particular measures are designed to prevent only \emph{a part} of Dragonblood's attacks, and does not affect one of them. Especially, the cache attack leveraging a password dependent control-flow of loop in the try-and-increment conversion function is neither discussed in this document, nor patched in most implementations (except for hostapd, which was the direct target of the original attack). We aim to raise awareness about this particular attack, and prove that we can extend it to gain additional information, with fewer measurements. To this end, we identify several implementations in which some code is executed only during the iteration where the password was correctly converted (or encoded). We show how an attacker can use cache attacks in order to leak some information on the password. We stress that the original Dragonblood attacks are still applicable on such implementations. However, our work takes a step further by leveraging some state-of-the-art techniques that improve the attack performance without changing the underlying threat model.

Indeed, we extend the original attack in which only the outcome of the first iteration is leaked. Using an unprivileged spyware, we demonstrate that attackers are able to learn the exact iteration where the first successful conversion occurred with high probability. We achieve this result by monitoring well-chosen memory-lines with a \textsc{Flush+Reload} attack~\cite{DBLP:conf/uss/YaromF14} to keep track of each iteration, and the success-specific code. We enhanced the reliability of our measurements by combing the attack to a Performance Degradation Attack (PDA)~\cite{DBLP:conf/acsac/AllanBFPY16}. Since the successful iteration is directly related to key exchange context (defined by both MAC addresses and the password), this leakage allows attackers to significantly reduce the number of measurements needed to recover the password. For instance, only 160 measurements are required in order to discard all the wrong passwords using the Rockyou dictionary~\cite{rockyou}, while Dragonblood needs 580 measurements. Roughly, we cut down the number of measurements by three, which makes our attack performs better in practice. 

We apply our findings on the wireless daemon iwd (iNet Wireless Daemon) that aims to replace wpa\_supplicant. Ironically, iwd is written by Intel and our identified vulnerabilities in their implementation are caused by Intel cache design. The version 1.0 was released in October 2019 (after the publication of Dragonblood) and it is already adopted by Arch Linux and Gentoo. We also extend our work to FreeRADIUS, which a widely deployed project used by millions of users~\footnote{\url{https://freeradius.org/about/\#usage\_statistics}}. We have not only communicated our findings to the maintainers of these two open-source projects, but also helped them to patch the vulnerable code.

The underlying technical details are quite similar concerning the identified vulnerability in iwd and FreeRADIUS. Therefore, for the sake of clarity and brevity, we will only detail the iwd case in the core of this paper. The FreeRADIUS case id discussed in Appendix~\ref{app:freeradius}) in order to highlight the specificity of their implementation. In summary, we make the following main contributions:
\begin{itemize}
\item We extended the original Dragonblood attack to recover not only the outcome of the first round, but the iteration yielding a successful conversion (see Section~\ref{sec:attack}).
\item We estimated the theoretical complexity of our attack and compared it to the original one (see Section~\ref{sub:attack_dictionary}).
\item We implemented a Proof of Concept of our attack, presenting practical results (see Section~\ref{sec:experimentation}).
\item We implemented mitigations and evaluated the overhead (see Section~\ref{subsec:countermeasures}).
\item We made all our code available\footnote{https://gitlab.inria.fr/ddealmei/poc-iwd-acsac2020/-/tree/master/}, from the testing environment setup using Docker, to the password recovery script.
\end{itemize}

Our attack illustrates the danger of overlooking a widely potential attack during a standardization process.
Therefore, we hope that our work would raise awareness concerning the need of constant-time algorithms by design that do not rely on savvy developers to provide secure implementations of ad-hoc mitigations.

\subsection{Attack Scenario}
We suppose a classical infrastructure where clients communicate with an access point (AP) across a wireless network. The goal of the attacker is to steal the password used to establish a secure communication with the AP. Once the password is compromised, the attacker can enter the network and perform malicious activities. 

In order to leverage the vulnerabilities defined in this paper, the attacker requires to perform two tasks. First, they need to install a spy daemon on a client station without any particular privilege. Second, they need to create a rogue AP that behaves as the legitimate AP, but can use different MAC addresses for different connections. 

Of course, we suppose that the rogue AP does not know the correct password, and therefore any session establishment between the rogue AP and a valid client will fail. Here, the goal of the rogue AP is to state different MAC addresses and to trick a client device to start a Dragonfly key exchange. Thus, the Wi-Fi daemon, using the correct password, will perform some operations that will be monitored by the attacker spy process. For each of these (failed) connections, the spy will generate a new trace that leaks the number of iterations needed to successfully encode the password. These bits of information are then used offline in order to prune a dictionary by verifying the number of iterations needed for each password. Each trace, with a different MAC address, yields a different iteration number. In our paper, we estimate that attackers require 16 traces to prune, for instance, the entire Rockyou dictionary. It is worth noting that, in our work, a trace generation needs 10 measurements with the same MAC address in order to guarantee a high accurate leakage.

\subsection{Responsible Disclosure}
Our attacks were performed on the most updated version of iwd and FreeRADIUS, as published at the time of discovery. We compiled both libraries using their default compilation flags, leaving all side-channel countermeasures in place. We reported our findings to the maintainers of iwd and FreeRADIUS following the practice of responsible disclosure. We further actively participated in coding as well as the empirical verification of the proposed countermeasures. Correspondingly, three patches were committed on the vulnerable projects: on iwd\footnote{\url{https://git.kernel.org/pub/scm/network/wireless/iwd.git/commit/?id=211f7dde6e87b4ab52430c983ed75b377f2e49f1}}, ell\footnote{\url{https://git.kernel.org/pub/scm/libs/ell/ell.git/commit/?id=47c2afeec967b83ac53b5d13e8f2dc737572567b}} (the underlying cryptographic library of iwd, also maintained by Intel), and FreeRadius\footnote{\url{https://github.com/FreeRADIUS/freeradius-server/commit/6f0e0aca4f4e614eea4ce10e226aed73ed4ab68b}}. On a side note, iwd maintainers prefered not to scrupulously respect the recommendations of the RFC 7664~\cite{rfc7664} by fixing the number of iterations to 30 (instead of 40). Moreover, we received special thanks from Alan Dekok, the project leader of FreeRADIUS, for our disclosure of the issue, and for helping with creating and verifying the fix. 

We did not issue any communication to the Wi-Fi Alliance, since the identified vulnerability is mainly caused by implementation flaws, and not the standard itself.

\section{Background} \label{sec:background}
%!TEX root = article.tex

In this section, we introduce the Dragonfly protocol, and describe the variant currently used in WPA3 and EAP-pwd.

\subsection{The Dragonfly Key Exchange} \label{sub:dragonfly}

Dragonfly is part of the Password Authenticated Key Exchange (PAKE) family. Its purpose is to use a low entropy password as an authentication medium, and to derive some high entropy cryptographic material from it. An important security requirement of PAKE protocols is to avoid offline dictionary attack: the only way an attacker should be able to get information about the password is to run the protocol with a guess and observe the outcome. Since Dragonfly is a \text{symmetric} PAKE, each party knows the password before initiating the protocol.

Dragonfly has been designed by Dan Harkins in 2008. In 2012, it has been submitted to the CFRG as a candidate standard for general internet use. This standardization ended up in 2015 by the release of RFC 7664~\cite{rfc7664}. Along with the protocol described in this standard, some other variants have been included in other protocols, such as TLS-pwd~\cite{rfc8492}, WPA3~\cite{7786995} or EAP-pwd~\cite{rfc5931}. These variants mainly differ by instantiation details, such as some constant values.

The security of Dragonfly is based on the discrete logarithm problem. Implementations can therefore rely on either Finite Field Cryptography (FFC) over multiplicative groups modulo a prime (MODP groups) or Elliptic Curve Cryptography (ECC) over prime field (using ECP groups). The exact workflow of the Dragonfly handshake varies slightly depending on the underlying group (ECP/MODP).
In order to avoid confusion, we adopt a classic elliptic curve notation: $G$ is the generator of a group, with order $q$. Lowercase denotes scalars and uppercase denotes group element. For elliptic curve, we assume the equation to be in the short Weirestrass form $y^2 = x^3 + ax + b \mod p$ where $a$, $b$ and $p$ are curve-dependent and $p$ is prime.

The protocol follows the same workflow for both side, meaning it can be performed simultaneously by both side, without attributing a role.
It can be broken down into three main parts: (i) password derivation; (ii) password commitment; and (iii) confirmation. 

Following the disclosure of \textit{Dragonblood} attack~\cite{DBLP:conf/sp/VanhoefR20} in 2019, both the Wi-Fi standard~\cite{IEEE_P802.11-19/1173r0} and EAP-pwd~\cite{draft_harkins_eap_pwd_prime_00} are updating the password derivation function of Dragonfly. Due to the fact that updates are long to be approved, and even longer to be deployed, current implementations of WPA3 still use the original derivation function, as described in~\cite{rfc7664}. In this section, we will focus on currently deployed implementations, hence the original design. 

\subsubsection{Password derivation} \label{ssub:dragonfly_pwd_derivation}

First, both the sender and the receiver need to convert the shared password into a group element. To do so, the standard describes a try-and-increment method called \emph{Hunting and Pecking}.
This approach consists in hashing the password along with the identity of both parties and a counter until the resulting value corresponds to a group element.
For MODP groups, this method, called hash-to-group, converts the password into an integer modulo $p$. For ECP groups, the method, called hash-to-curve, converts the password into the x-coordinates of an elliptic curve point. The y-coordinate is chosen at the end from the parity of the digest. The pseudocode describing this process on ECP groups is summed-up in Listing~1.

%%\ref{listing:hunt_and_peck}

\begin{center}
\noindent \includegraphics[width=\columnwidth]{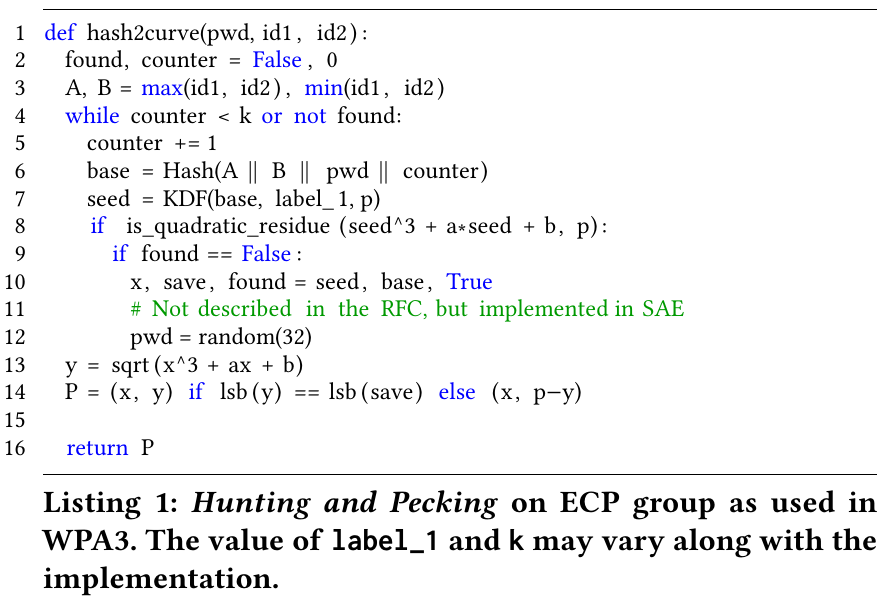}
\end{center}
%%%\begin{lstlisting}[
%%%caption={\emph{Hunting and Pecking} on ECP group as used in WPA3. The value of \texttt{label\_1} and \texttt{k} may vary along with the implementation.},
%%%label={listing:hunt_and_peck},
%%%float,floatplacement=H,mathescape=true]
%%%def hash2curve(pwd, id1, id2):
%%%	found, counter = False, 0
%%%	A, B = max(id1, id2), min(id1, id2)
%%%	while counter < k or not found:
%%%		counter += 1
%%%		base = Hash(A || B || pwd || counter)
%%%		seed = KDF(base, label_$1$, p)
%%%		if is_quadratic_residue(seed^3 + a*seed + b, p):
%%%			if found == False:
%%%				x, save, found = seed, base, True
%%%				# Not described in the RFC, but implemented in SAE
%%%				pwd = random(32) 
%%%	y = sqrt(x^3 + ax + b)
%%%	P = (x, y) if lsb(y) == lsb(save) else (x, p-y)
%%%	
%%%	return P
%%%\end{lstlisting}

Along the standardization process, various design flaws have been identified regarding the {password-dependent} nature of this function. Therefore, some mitigations were introduced to avoid password-dependent time variation in the execution of the function. Indeed, the number of rounds needed to find a value $x$ that corresponds to a point on the curve is directly related to the password and the parties identities. First, the standard mandates a fixed number of iterations in the derivation loop, noted $k$, regardless of the correct iteration. Setting this limit at $k = 40$ is recommended to minimize the risk of a password needing more iterations. All extra operations are performed on a random string, with no impact on the resulting element. Generating a dummy string for the extra operations is not described in RFC 7664, but has been discussed by the CFRG during the standardization process, and has been included in deployed variants of Dragonfly (such as TLS-pwd~\cite{rfc8492} and SAE~\cite{7786995}). In our paper, we show that such an operation is not enough to defend against our cache attacks.

Sensitive information may also leak when checking for the validity of the potential x-coordinate (Listing~1, line 7). Indeed, WPA3 mandates to compute the Legendre before computing $y$. However, textbook Legendre may not be constant time and leak information about the value of $x$~\cite{DBLP:conf/crypto/Icart09}. To overcome this issue, the protocol has been updated~\cite{IEEE_P802.11-14/0640r1,IETF:mail_archive/cfrg/WXyM6pHDjGRZXZzSc_HlERnp0Iw} to blind the computations by generating a random number for each test, squaring it, and multiplying it to the number being tested. The result is then multiplied by a per-session random quadratic (non-)residue before computing the Legendre symbol. The square root is then computed once and for all at the end of the function.

\subsubsection{Commitment and Confirmation phase}
\begin{figure}
	\includegraphics[width=\linewidth]{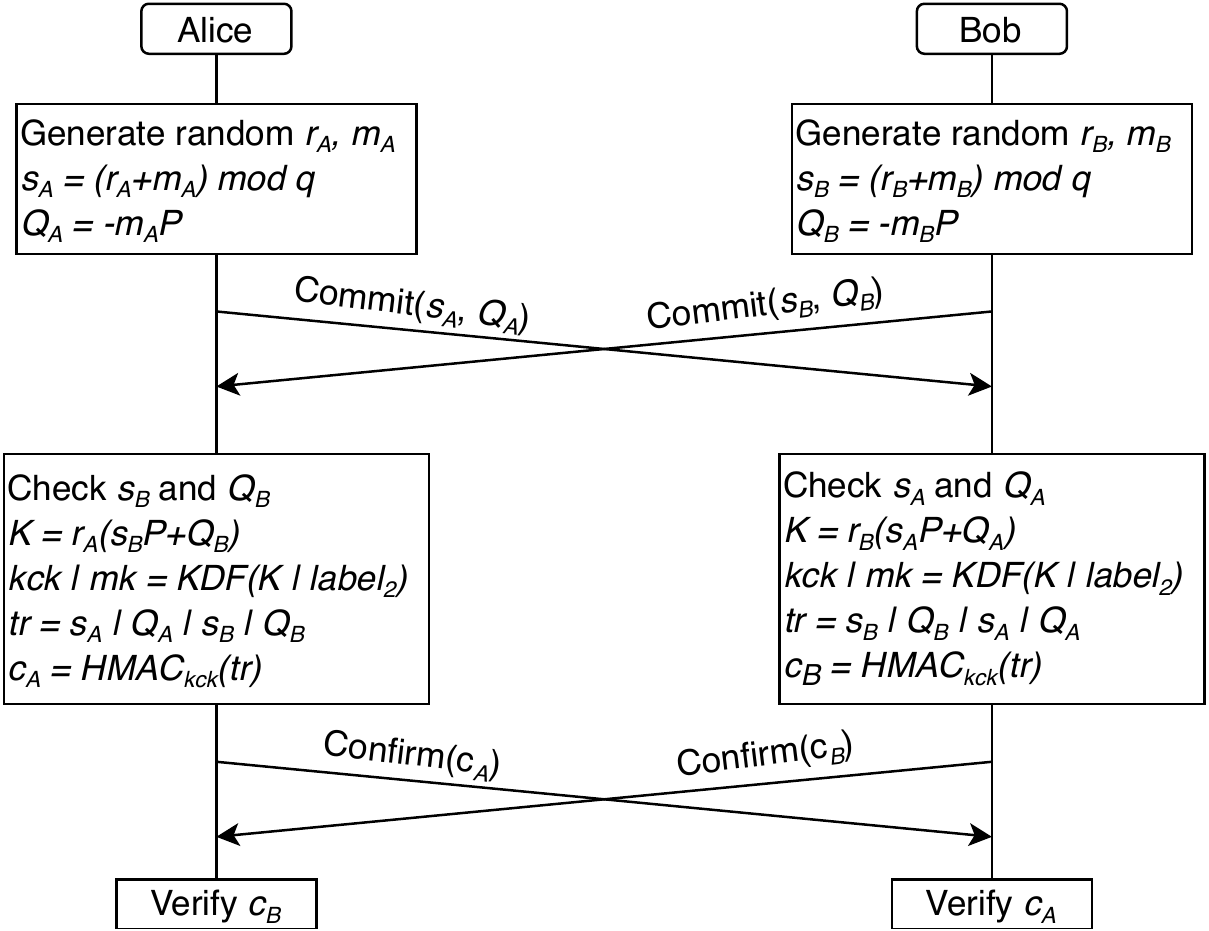}
	\caption{Dragonfly handshake workflow. $P$ is the group element derived from the password, and $label_2$ is a string that may vary along with the protocol in which the handshake is performed.}
	\label{fig:dragonfly_workflow}
\end{figure}

Once the shared group element has been computed, both parties exchange a commit frame followed by a confirmation frame to conclude the handshake, as illustrated in Figure~\ref{fig:dragonfly_workflow}. 

The commit frame is built with two values: a commit scalar $s_i = r_i + m_i \mod q$, computed by adding two random numbers $r_i,\ m_i \in [2, q)$, and a commit element $Q_i = -m_iP$. 
When receiving this frame, a party needs to check if the value $s_i$ is in the bounds (\textit{i.e.} $s_i \in [2, q)$) and if the commit element $Q_i$ belongs to the group. A failure in any check results in aborting the handshake.

In the confirmation phase, both parties compute the master key $K$. For MODP groups, the key can be used as is, but the x-coordinate is extracted in case of ECP group. This value is then derived into two sub keys using a KDF: $kck$ is a confirmation key and $mk$ is used as a master key for the subsequent exchanges.
Using the confirmation key, HMAC is computed over the transcript of the session. The resulting tag is included then in a confirm frame, to be verified by the other party. The handshake succeeds only if both verification ends successfully.

\subsection{Integration of Dragonfly in WPA3}

WPA3 uses a slight variant of Dragonfly, called Simultaneous Authentication of Equals (SAE)~\cite{7786995}. In this particular variant, the label values are fixed and each party is identified by its MAC address ($id1$ and $id2$ in Listing~1).

The SAE handshake is executed between the client and the access point (AP) in order to compute the Pairwise Master Key (PMK), called $mk$ in Figure~\ref{fig:dragonfly_workflow}. Afterward, a classic WPA2 4-way handshake is performed with this PMK  in order to derive fresh cryptographic material.
Since the entropy of the initial master key is significantly higher than in WPA2, the dictionary attack on the 4-way handshake is no longer relevant.

\subsection{Micro-architectural Preliminaries} 

\subsubsection{Cache architecture}
To mitigate the gap between slow memory access and fast processing, CPU benefits from fast access caches that are located close to the processor cores. The storage capacity is kept small, so only currently or recently used data are stored.
On modern processors, the CPU cache is usually divided into several levels following an access hierarchy. Higher-level caches are closer to the core and typically smaller and faster than lower-level caches.
In classical Intel architecture, which we will consider from now on, CPU cache is divided into three levels.
Each core has two dedicated caches, L1 and L2, shared by all processes executing on the same core. The third cache, called Last-Level-Cache (LLC) is shared between all cores, hence all the processes.

When the CPU tries to access a memory address, it will first check the higher level cache. If the memory line has been cached, the processor will find it (\textit{cache hit}). Otherwise, in a \textit{cache miss}, the processor will keep looking in lower memory, down to the DRAM if needed. Once the appropriate memory line is found, the processor saves its content in cache for a faster access in the near future.

Finally, in modern Intel CPUs, the LLC has a significant property of being inclusive, meaning that it behaves as a superset of all higher caches. An important consequence of this feature, exploited in some attacks, evicting a memory line from the LLC will also have impacts on L1 and L2 caches.

\subsubsection{Cache optimizations}	\label{ssub:cache_optimizations}
In some cases, memory lines can be brought to cache even though they are not accessed. This is due to some cache optimization, that makes the exact cache behavior difficult to predict. For instance, Intel's prefetcher (\cite{IntelArchitectureOptimizationManual}, Chapter 7) will pair consecutive memory lines and attempt to fetch the pair of a missed line to avoid looking for it in the near future. It may also detect memory access patterns and prefetch the lines to be loaded next.

\subsubsection{Micro-architectural leaks}
The time taken to access some data will significantly change whether the data is already in a CPU cache (cache hit), or if the CPU needs to look for it in the RAM (cache miss). This cache interaction can be triggered by two behaviors: (i) the CPU needs to access some data;
(ii) the CPU needs to access some instruction.

In both cases, this can lead to a vulnerability if the element to access is related to some secret information (e.g. the index of the array or the instruction to access depends on a secret value).
Given this information, an attacker can use a spying process interacting in a particular way with the cache to trigger different timing of memory access. The nature of the interaction defines various types of attacks, each having benefits and drawbacks. Most instruction-driven attacks consist in probing the victim code, and inferring some data from the instructions performed.

Depending on the threat model and the targeted architecture, an attacker may or may not be able to access low level caches shared between two threads. However, the LLC is shared between all cores. From now on, \textit{cache} will refer to the LLC unless specified otherwise.

\subsection{Related Work} \label{sub:related_work}
Micro-architectural attacks have long been used to gain information about sensitive data. In 2014, Yarom and  Falkner~\cite{DBLP:conf/uss/YaromF14} presented a ground breaking approach called \textsc{Flush+Reload}. Unlike previous approaches, which infer victim memory line access based on the cache set activity, the novel approach directly monitors memory access in the inclusive L3 cache, yielding more interesting results. Since then, this method has been exploited to recover sensitive information in various contexts~\cite{DBLP:journals/iacr/YaromB14,DBLP:conf/ches/BengerPSY14,DBLP:conf/ctrsa/PolSY15,DBLP:conf/ches/BruinderinkHLY16,DBLP:conf/ches/YaromGH16,DBLP:conf/acsac/AllanBFPY16,DBLP:conf/ccs/PesslBY17, DBLP:conf/ccs/GenkinVY17, DBLP:journals/tches/AldayaGTB19, DBLP:conf/sp/CohneyKPGHRY20, DBLP:journals/iacr/AranhaNTTY20}.

In 2016, Allan et al. enhanced the leakage by introducing the Performance Degradation Attack~\cite{DBLP:conf/acsac/AllanBFPY16}. The goal is to systematically evict some well chosen memory line in order to make the leakage easier and more reliable to exploit.

The Dragonfly handshake has already been reviewed in the past. A first version was found vulnerable to offline dictionary attack~\cite{IETF:mail_archive/cfrg/mGnSNL8QW_fuCTwcyvh8lY9Z5G0}. In 2014, Clarke and Hao outlined a small subgroup attack due to a lack of verification by the parties~\cite{DBLP:journals/iet-ifs/ClarkeH14}.
In 2019, Vahoef and Ronen identified several flaws in different implementations of Dragonfly, namely in WPA3 and EAP-pwd~\cite{DBLP:conf/sp/VanhoefR20}. They outlined various vulnerabilities at the protocol level as long as at the implementation level. They demonstrated that some implementations of the hash-to-curve method leak sensitive information through micro-architectural attacks. Exploiting these leaks with a classic \textsc{Flush+Reload} attack, they were able to learn the outcome of the first quadratic residue computation, and therefore they could learn if the password was successfully derived at the first iteration or not.
We go one step further and demonstrate that combining \textsc{Flush+Reload} and a well chosen PDA, we are able to learn the exact iteration corresponding to the successful derivation, which allows us to increase the probability of success, while significantly decreasing the complexity of the attack with fewer traces and computations.

Tschacher Master thesis~\cite{Tschacher:2019} offers valuable insight on how test environment for WPA3 protocol fuzzing shall be implemented.

\section{Attacking iwd implementation} \label{sec:attack}
%!TEX root = article.tex

In this section, we extend the cache-based attack presented by Vahoef and Ronen in 
Dragonblood~\cite{DBLP:conf/sp/VanhoefR20}. Indeed, the attack of~\cite{DBLP:conf/sp/VanhoefR20} (in Section 6)
allows attackers to only learn the outcome of the first derivation attempt, and needs a high number of traces with 
different MAC addresses to be effective. Thus, various WPA3 implementations have just decided to overlook such an attack, and rather prioritize patching other vulnerabilities~\cite{radiator_security_advisory}.

In our attack, we greatly reduce the required traces by exactly estimating 
the number of iterations for a particular password with high probability. Then, we show how our attack can be used 
to guess the target password by tremendously cutting down the dictionary size.

We demonstrated our attack on iNet Wireless Daemon\footnote{https://git.kernel.org/pub/scm/network/wireless/iwd.git/} 
(iwd) version 1.8 (current version as of the time of writing), but we believe that our work is applicable to any unpatched 
implementation that is still vulnerable to the initial cache-attack (see Appendix~\ref{app:freeradius} for the case of the current version of FreeRADIUS).

\subsection{Threat Model} \label{sub:threat_model}
Our attack targets Wi-Fi network, either a client or an Access Point (AP). Thus, we assume that the attacker to be within 
range of the physical target. To efficiently reduce the set of potential passwords, attackers need to monitor multiple 
handshakes, involving the same password and different MAC addresses. When the target is an AP, this can easily be 
done either by waiting for a client to connect, or by playing the role of a client. 
If attackers target clients, they can setup multiple rogue clones of the legitimate AP, advertizing stronger signal strength 
(thereby making the client automatically choosing it) and different MAC addresses. If clients are already connected to 
the legitimate AP, attackers can force a de-authentication beforehand~\cite{DBLP:conf/uss/BellardoS03,DBLP:conf/acsac/VanhoefP14}. 
Blocklist mechanisms are usually limited, since implementations tend to apply them based on the MAC address of 
the AP (that can easily be forged). We note that iwd might automatically generate a new random MAC address every 
time the daemon starts (or if an interface is detected, due to a hot-plug for instance). However, the default configuration 
uses one permanent address. We note that using different MAC addresses is not relevant to EAP-pwd, that is the Dragonfly variant in FreeRADIUS (see Appendix~\ref{app:freeradius} for further details).

Due to the micro-architectural nature of the leak, attackers need to be able to monitor the CPU cache, using a classical
 \textsc{Flush+Reload} attack for instance. Since cache access and eviction do not rely on particular permissions, the 
 most common assumption is that attackers can deploy an unprivileged user-mode program in the targeted device. 
 This spy process runs as a background task and records the CPU cache access to some specific functions.
Papers in the literature also suggest that such memory access can be granted remotely, performing the attack through 
JavaScript code injection in web browser~\cite{DBLP:conf/ccs/OrenKSK15}. However, we did not investigate the effectiveness 
of our attack in such a context.

\subsection{IWD Implementation}
The Dragonfly exchange implemented in iwd follows the standard SAE~\cite{7786995}. Only the ECP-groups variant is 
supported with the NIST's curves P256 and P384. The corresponding \emph{Hunting and Pecking} is implemented in the 
function \texttt{sae\_compute\_pwe}, as illustrated in Listing~2.

\begin{center}
	\noindent \includegraphics[width=\columnwidth]{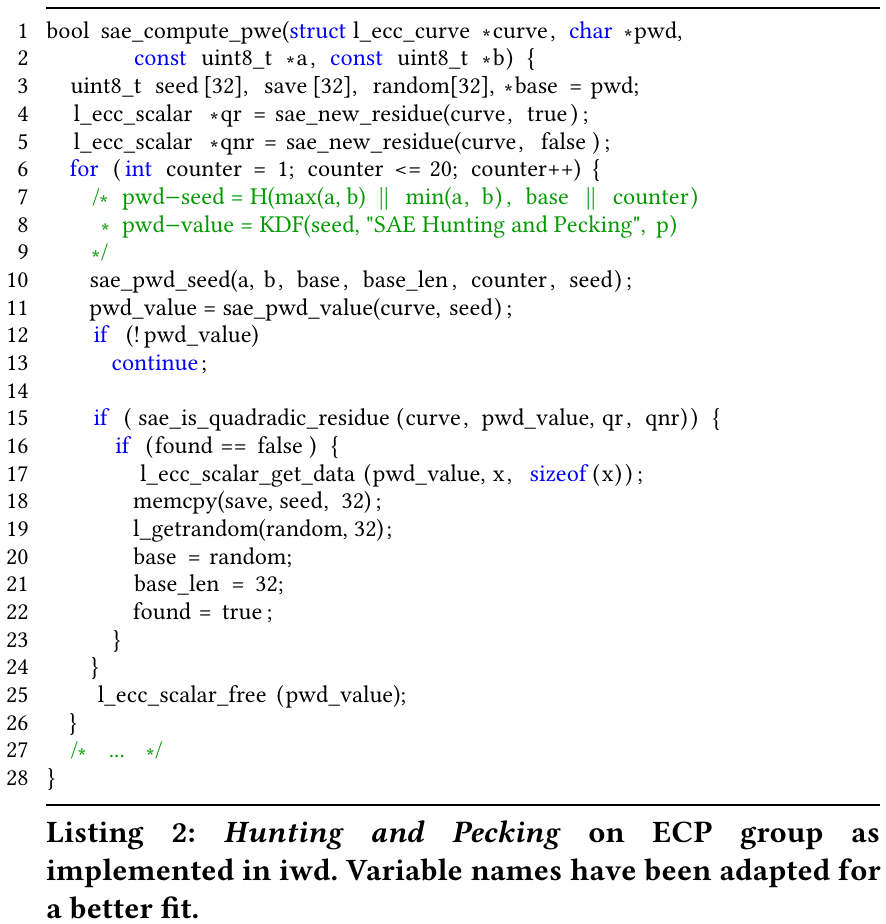}
\end{center}

Each type or function starting by \texttt{l\_*} refers to a function in the Embedded Linux 
Library\footnote{https://git.kernel.org/pub/scm/libs/ell/ell.git/} (ell), a minimalist cryptographic library developed by Intel.
By default, this library is statically linked to the binary at compilation time. Users can decide to use a dynamic linking by 
specifying the correct option before compiling. We stress that the linking strategy does not impact the result of our attack; 
only some details in the addresses to monitor are concerned (see Section~\ref{sub:attack_detail}).

%%%\begin{lstlisting}[
%%%	caption={\emph{Hunting and Pecking} on ECP group as implemented in iwd. Variable names have been adapted for a better fit.},
%%%	label={listing:iwd_compute_pwe},
%%%	language=C,
%%%	float,floatplacement=H]
%%%bool sae_compute_pwe(struct l_ecc_curve *curve, char *pwd,
%%%				const uint8_t *a, const uint8_t *b) {
%%%	uint8_t seed[32], save[32], random[32], *base = pwd;
%%%	l_ecc_scalar *qr = sae_new_residue(curve, true);
%%%	l_ecc_scalar *qnr = sae_new_residue(curve, false);
%%%	for (int counter = 1; counter <= 20; counter++) {
%%%		/* pwd-seed = H(max(a, b) || min(a, b), base || counter)
%%%		 * pwd-value = KDF(seed, "SAE Hunting and Pecking", p)
%%%		*/
%%%		sae_pwd_seed(a, b, base, base_len, counter, seed);
%%%		pwd_value = sae_pwd_value(curve, seed);
%%%		if (!pwd_value)
%%%			continue;
%%%		
%%%		if (sae_is_quadradic_residue(curve, pwd_value, qr, qnr)) {
%%%			if (found == false) {
%%%				l_ecc_scalar_get_data(pwd_value, x, sizeof(x));
%%%				memcpy(save, seed, 32);
%%%				l_getrandom(random, 32);
%%%				base = random;
%%%				base_len = 32;
%%%				found = true;
%%%			}
%%%		}
%%%		l_ecc_scalar_free(pwd_value);
%%%	}
%%%	/* ... */
%%%}
%%%\end{lstlisting}

It is easy to notice that explicit branching at lines 15 and 16 makes the control flow input-dependent.  An attacker who is able 
to tell at what iteration the code between line 17 and 22 is executed can guess how many rounds are needed before successfully 
returning from this function.

\subsection{Cache-Attack Details}	\label{sub:attack_detail}
In order to efficiently determine at what iteration a password is successfully converted, the attackers' needs are twofold. First, 
they need to be able to distinguish each iteration. Second, they shall guess when the success-specific code (lines 17-22) is executed. 

To achieve the first goal, we create a synchronization clock by monitoring some memory line accessed at the beginning of each 
loop. The call to \texttt{kdf\_sha256}, a function of \texttt{libell} called inside \texttt{sae\_pwd\_value}, is a good candidate. More specifically, 
we monitor a memory address corresponding to the loop calling this hash function. Thanks to the complex nature of this operation, 
we were able to detect access to this call every time. Moreover, this particular memory address is not accessed during the rest of 
the protocol, thereby avoiding any potential noise in our traces. 

Monitoring access to the code executed on success is less straightforward: the address range to be accessed inside 
\texttt{sae\_compute\_pwe} is too small and too close to the rest of the loop to be reliably monitored. The best choice is to monitor 
instruction in one of the functions called at lines 17 to 19. Tests have shown that monitoring inside \texttt{l\_getrandom} yields the 
best results: other functions are called too often, at various places, bringing noise to our traces. However, random number generation 
is also part of the quadratic residue verification (\texttt{sae\_is\_quadratic\_residue}, line 15) in order to blind the computation. 
Fortunately, these accesses can be distinguished given the number of cycles elapsed since the beginning of the iteration.

Due to complex CPU optimization techniques (see Section~\ref{ssub:cache_optimizations}) and some system noise, the measurements are noisy and some traces may yield incorrect results. Moreover, a call to \texttt{l\_getrandom} is usually performed in a few cycles, implying that we can miss this call due to the temporal resolution of \textsc{Flush+Reload}.

In order to significantly improve the reliability of our results, we combined the \textsc{Flush+Reload} attack with the Performance Degradation attack (PDA), as presented in~\cite{DBLP:conf/acsac/AllanBFPY16}. Since the first call to \texttt{l\_getrandom} occurs before the proper quadratic residue check, we evict a memory line inside the code in charge of the Legendre symbol computation. Hence, we significantly increase the delay between our synchronization clock and the success-specific code, while keeping a low delay to reach the first call to \texttt{l\_getrandom}.

To sum up, by simply monitoring two addresses with a classic \textsc{Flush+Reload} technique, and repeatedly evicting a single memory address, we were able to collect traces that yield more relevant results with only a few samples.

\subsection{Miscellaneous Leak}
As specified in the Dragonfly RFC~\cite{rfc7664} and in the SAE standard~\cite{7786995}, the number of iterations to perform during the password conversion is not fixed. It can be defined as any non-negative integer, providing it effectively guarantees a successful conversion with high probability. RFC 7664 advises to set $k$ to at least 40 in order to get roughly one password over one trillion that needs more iterations.

As for iwd, the implementation sets $k = 20$, making this probability significantly lower, with about one over $2\cdot 10^6$ passwords requiring more than $k$ iterations. In practice, using only password drawn from existing dictionaries~\cite{rockyou,crackstation:HumanPassword}, we were able to find a consistent list of password needing more than 20 iterations (see Appendix~\ref{app:password_2X_iter} for a sample). Using these password related dictionaries, with random MAC addresses, we found an average of 33.6 passwords ($9.5\cdot 10^{-5} \%$ of the dictionaries).

In this scenario, a client would be unable to authenticate to the AP until the password or the MAC address of one party is changed.
From an attacker perspective, finding such a tuple provides a lot of information on the password, without the aforementioned cache-attack. Indeed, they can assume that the password needs at least 20 iterations, and perform an offline dictionary attack as described in Section~\ref{sub:attack_dictionary}. However, due to the low probability of finding these tuples, we did not take it into account in the rest of the paper.

\subsection{Dictionary Partitioning}	\label{sub:attack_dictionary}
By exploiting the leakage presented above, attackers can significantly reduce the set of potential passwords with an offline brute-forcing program. Given a dictionary and some $m$ collected traces, it iterates over the passwords and eliminates those that do not yield the same result when derived with the corresponding MAC addresses. The remaining passwords, giving the same results, are potential candidates that now constitute the new dictionary.

\subsubsection{Theoretical success rate}
Let each leak be represented by a tuple $(A,\ B,\ k)$ with $A$, $B$ the MAC addresses and $k \in [1,\ 20]$ the number of iterations. 
When converting a password into a group element, the success of each iteration is bounded to the success of the quadratic residue test. Let be $p$ the order of the underlying field and $q$ the order of the generator? Since Dragonfly only support elliptic curves of cofactor $h=1$, $q$ also denotes the total number of points on the curve. Then, a random integer $x\in[0, p)$ is a quadratic residue with probability: 
\begin{equation}
	p_s = \frac{q}{2p} \approx 0.5 \approx 1 - p_s.
\end{equation}

The input of the quadratic residue is considered random (being the output of a KDF). Hence, each iteration is independent of the others if we model 
the KDF as a random oracle. Let $X$ denote the random variable representing the number of iterations of a trace, and $k \in [1,\ 20]$:
\begin{equation}
	\Pr[X = k] = p_s^k.
\end{equation}

The probability for a trace to eliminate any tested password depends on the number of iterations $k$. Let $Y_1$ be the random variable representing the success (1) or the failure (0) of a password to pass each test in a single trace. We got $Y_1 = 1$ only if the password succeeds all tests, \textit{i.e.} with probability $\Pr[X = k]$, hence:
\begin{equation}
	\Pr[Y_1 = 0\ |\ X = k] = 1 - \Pr[X = k] = 1 - p_s^k.
\end{equation}
More generally, the probability for a password to be eliminated by a random trace is:
\begin{equation}
	\Pr[Y_1 = 0] = \sum_{i=1}^{20} \Pr[X = i] \cdot \Pr[Y_1 = 0 | X = i].
\end{equation}

Hence, the probability for a password to be pruned by at most $n$ traces is the sum of probabilities for it be pruned either at the first trace or to pass the first and be pruned at the second, and so forth:
\begin{equation}
	p_{y_n} = \Pr[Y_n = 0] = \sum_{i=0}^{n-1} \Pr[Y_1 = 0]\cdot (1-\Pr[Y_1 = 0])^i.
\end{equation}

Let $L$ be the size of our dictionary, and $d$ be the number of passwords we want to eliminate. Let $Z_n$ be the number of passwords we remove using $n$ traces. Since tests behave as independent trials, $Z_n$ follows a binomial law, hence:
\begin{equation}
	\Pr[Z_n \geq d] = \sum_{i=d}^{L}{L\choose i} \cdot p_{y_n}^i \cdot (1-p_{y_n})^{L-i}.
\end{equation}

Table~\ref{tab:comp_traces} gives an overview of the number of traces required to eliminate all wrong passwords from different dictionaries, with a probability greater than 0.95. We outline the benefit of our attack compared to the original Dragonblood's, reducing the average number of required traces by roughly 43\%.
In practice, we do not need to remove all passwords from the dictionary, we only need to reduce it enough, so that remaining passwords can be tested in an active attack. Keeping more passwords in the dictionary would reduce the number of required traces.

\begin{table}[]
	\centering
\begin{imageonly}
	\begin{tabular}{l|r|l|r|l|}
		\cline{2-5}
		& \multicolumn{2}{r|}{\textbf{Dict. size}}                       & \textbf{Avg traces} & \textbf{\begin{tabular}[c]{@{}l@{}}Avg traces \\ in \cite{DBLP:conf/sp/VanhoefR20}\end{tabular}} \\ \hline
		\multicolumn{1}{|l|}{Rockyou}                   & \multicolumn{2}{r|}{$1.4 \cdot 10^7$}  &            16             & 29                                     \\ \hline
		\multicolumn{1}{|l|}{CrackStation} & \multicolumn{2}{r|}{$3.5 \cdot 10^7$}  &            17             & 30                                     \\ \hline
		\multicolumn{1}{|l|}{HaveIBeenPwned}            & \multicolumn{2}{r|}{$5.5 \cdot 10^8$}  &             20            & 34                                     \\ \hline
		\multicolumn{1}{|l|}{8 characters}    & \multicolumn{2}{r|}{$4.6 \cdot 10^{14}$} &           32              & 53                                     \\ \hline
	\end{tabular}
\end{imageonly}
	\caption{A Comparison of the Number of the Required Traces to Prune all Wrong Passwords Between Our attack and Dragonblood.}
	\label{tab:comp_traces}
\end{table}

\subsubsection{Complexity of the offline search}

Each test we perform is bounded by the complexity of a quadratic residue test (which is basically a modular exponentiation). The theoretical cost of such an operation has already been discussed in~\cite{DBLP:conf/sp/VanhoefR20}, and can be applied the same way in our context. Authors estimated, given their benchmark of the \texttt{PowMod} function~\cite{XMPlib} on an NVIDIA V100 GPU, that approximately $7.87\cdot10^9$ passwords per second can be tested.
Since each test is independent, the amount of parallelization is up to the attacker capacity, and can be higher. Namely, one can choose to split the dictionary into $k$ pieces and run $k$ instances of the dictionary reducer.

\section{Experimental results} \label{sec:experimentation}
%!TEX root = article.tex

In this section, we describe our setup and give details about the experimental results we obtained during our evaluation. 
All the scripts and programs we used are made open-source\footnote{https://gitlab.inria.fr/ddealmei/poc-iwd-acsac2020/-/tree/master/}.

\subsection{Experimental Setup} \label{ssec:setup}
Our experiments were performed on a Dell XPS13 7390 running on Fedora 31, kernel 5.6.15, with an Intel(R) Core(TM) 
i7-10510U and 16 GB of RAM. Binaries were compiled with gcc version 9.3.1 build 20200408 using the default configuration 
(optimization included). Namely, the Embedded Linux Library version 0.31 was statically linked to iwd during compilation.

During our experiment, we deployed hostapd (version 2.9) as an Access Point, and iwd (version 1.7) as a client. Both were 
installed and launched on the same physical device, using emulated network interfaces, as described in~\cite{Tschacher:2019}.

We kept the default configuration on both ends, meaning the key exchange is always performed using IKE group 19, 
corresponding to P256. Similar results would have been observed using group 20 (curve P384) by tweaking the threshold 
of our spy process. 

Our spy process has been implemented by following classical \textsc{Flush+Reload} methods. Moreover, we used Mastik 
v0.02 implementation of the PDA~\cite{Mastik}.

\subsection{Trace Collection}
Once both client and AP were setup to use a password that was randomly drawn from a dictionary, we launched the spy 
process to monitor well-chosen memory lines (see Section~\ref{sub:attack_detail}). After each connection, we disconnected 
the client and reconnected it a few times to acquire multiple samples. This step emulates a de-authentication attack aiming 
at collecting multiple samples with the same MAC addresses. For each password we went through this process using 10 
different MAC addresses, allowing us to acquire up to 10 independent traces for the same password. For each MAC address, 
we collect 15 samples. Our observations were consistently obtained through testing 80 passwords in order to evaluate the 
effectiveness and the reliability of our trace collection techniques.

We call \emph{sample}  the result of monitoring one Dragonfly key exchange, with a fixed password and MAC addresses. 
It is represented by succession of lines, corresponding to either a call to the synchronization clock (\texttt{kdf\_sha256}) or 
to \texttt{l\_getrandom}. The value following each label is an indicator of the delay since the last call to the synchronization 
clock. An example can be found in Appendix~\ref{app:trace_sample}, corresponding to a trace yielding four iterations.
A trace is a collection of samples, all corresponding to the same password and the same MAC address.

\subsection{Trace Interpretation}

We also designed a script that automatically interprets our traces and outputs the most probable iteration in which the 
process of password conversion first succeeds.

The trace parser process is described in Listing~3. The core idea is to first reduce the noise by 
eliminating all poorly formed samples (which could not be interpreted anyway, often because of system noise). Then, 
each sample is processed independently, contributing to the creation of a global trace score. To do so, each line of a 
sample is read, and depending on the corresponding label, it is processed as follow: (i) if the label is the synchronization 
clock, we increase the iteration counter by one; (ii) otherwise, the score of the current iteration is increased by the delay 
associated to that line. In the latter case, if the delay is long enough (the threshold may be architecture specific), we can 
stop the parsing of that sample and process the next one. Once every sample of a trace has been processed, the 
score of each iteration comes at as indicator of the most probable successful iteration.

Since false positives have severe consequences, we chose to eliminate any trace that does not yield a clear result. 
In such a case, the script raises a warning to the attacker for future manual interpretation.

\begin{center}
	\noindent \includegraphics[width=\columnwidth]{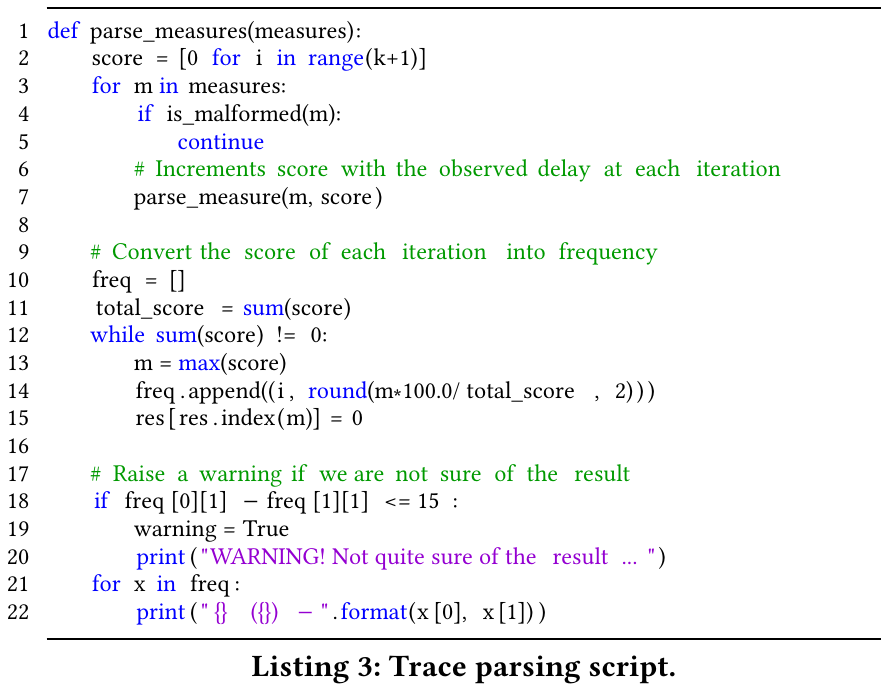}
\end{center}

%\begin{lstlisting}[
%	caption={Trace parsing script.},
%	label={listing:trace_parser},
%	language=Python,
%	tabsize=4,
%	float,floatplacement=H]
%def parse_measures(measures):
%	score = [0 for i in range(k+1)]
%	for m in measures:
%		if is_malformed(m):
%			continue
%		# Increments score with the observed delay at each iteration
%		parse_measure(m, score)
%
%	# Convert the score of each iteration into frequency
%	freq = []
%	total_score = sum(score)
%	while sum(score) != 0:
%		m = max(score)
%		freq.append((i, round(m*100.0/total_score , 2)))
%		res[res.index(m)] = 0
%
%	# Raise a warning if we are not sure of the result
%	if freq[0][1] - freq[1][1] <= 15 :
%		warning = True
%		print("WARNING! Not quite sure of the result...")
%	for x in freq:
%		print("{} ({}) - ".format(x[0], x[1]))
%\end{lstlisting}

\subsection{Results}

\begin{figure}
	\includegraphics[width=\linewidth]{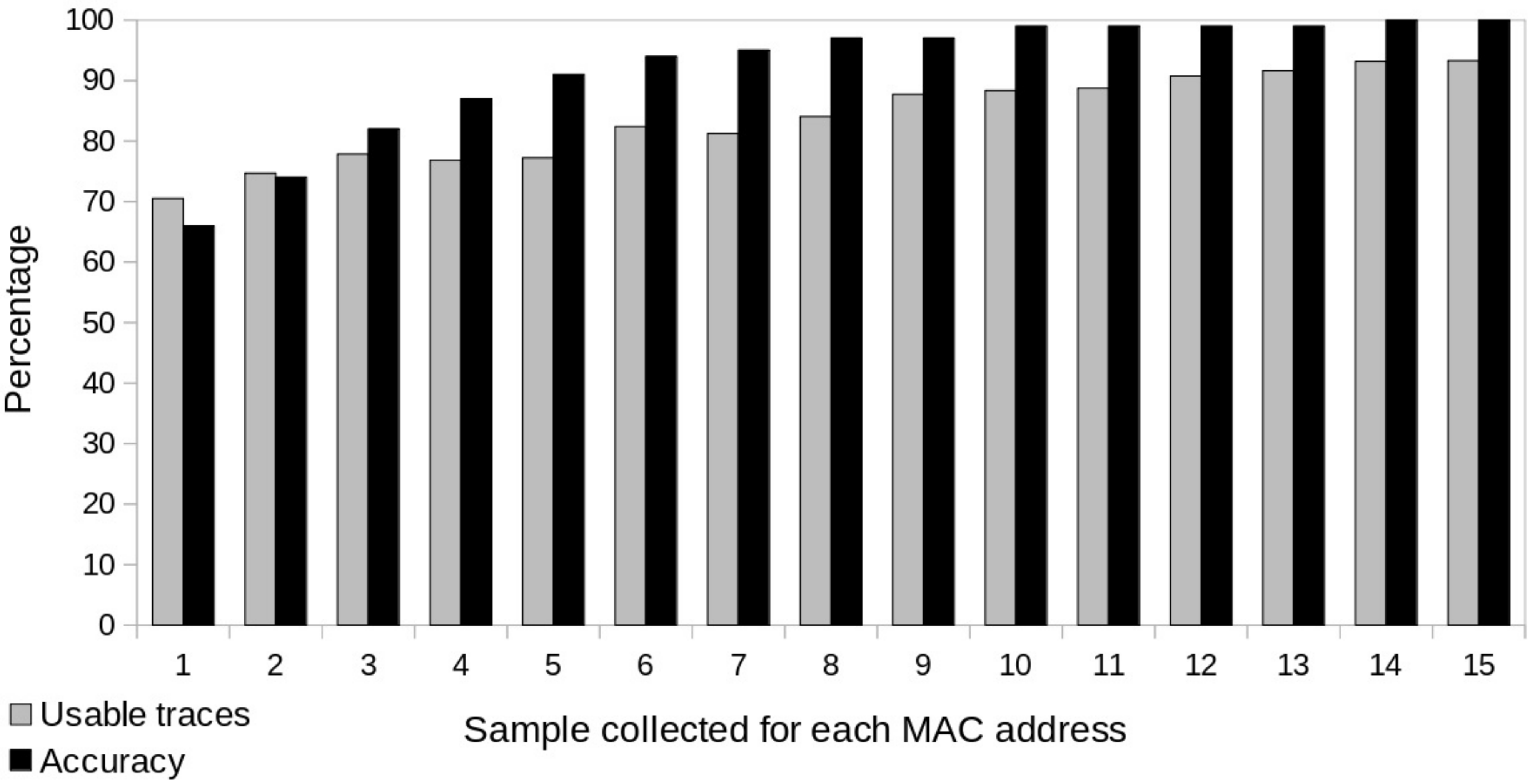}
	\caption{Reliability of our experiment given a different number of samples to interpret for each MAC address. 
	Accuracy represents the closeness of our prediction to the real value. Usable traces represent the percentage 
	of traces we were able to automatically exploit, without high risk of miss-prediction.}
	\label{fig:experimentation_results}
\end{figure}

We summed-up the results of our experimentations, with different number of samples for each MAC address, 
in Figure~\ref{fig:experimentation_results}. With only one measurement per address, approximately 70.5\% of 
the traces can be automatically interpreted (others have a high risk of miss-prediction). However, the accuracy 
of our prediction is only 66\%. We need to collect 5 samples to achieve an accuracy greater than 90\% (with 
77\% of usable traces). We achieve 99\% accuracy with only 10 measurements, with a trace usability of 88\%.

We stress that trace usability only represents the ability for the \emph{parser} to \emph{automatically} interpret 
the trace. For most warnings, a manual reading of the samples (about 1-2 minutes) allows attackers to successfully 
predict the round (some measurements do not yield a clear result, and should be ignored). We also note that 
even if our script was unable to decide between two adjacent values, \textit{e.g.} five and six, we can assume 
that more than four iterations are required for password conversion.

These results outline the improvement of our attack compared to Dragonblood. In~\cite{DBLP:conf/sp/VanhoefR20}, 
at least 20 samples were needed for each MAC address to achieve a success rate of 99\% (only 10 in our attack). 
Moreover, with our attack, each successfully interpreted trace gives at least as much information, and roughly twice 
more on average (see Section~\ref{sub:attack_dictionary}). Consequently, our work greatly reduces the number of 
the required measurements (or samples) in order to prune all wrong passwords in a given dictionary. For instance, 
our work needs 160 measurements for the Rockyou dictionary, while Dragonblood needs 580 measurements. Roughly 
speaking, the measurements are cut down by at least 3. Moreover, our attack requires to vary the MAC addresses less often (almost twice as fewer). Thus, our work performs better in practice, particularly in a context where cache-based measurements are limited. Of course, we 
argue that our results can be generalized for other implementations suffering from the same type of vulnerability.

\section{Discussion and Conclusion} \label{sec:conclusion}
%!TEX root = article.tex

%\begin{lstlisting}[
%caption={Python-like pseudocode of a constant time version of \emph{Hunting and Pecking} on P256.},
%label={listing:branch_free_patch},
%language=Python,
%float,floatplacement=H]
%# Constant time binary buffer selection copy. All operations have 
%# identical memory access pattern
%def const_time_select(mask, true_val, false_val, dst):
%	for i in range(len(dst):)
%		dst[i] = (mask & true_val) | (~mask & false_val)
%
%# Hunting and Pecking function
%def sae_compute_pwe(curve, pwd, addr1, addr2) {
%	x, x_cand = bytearray(32), bytearray(32)
%	save, found = 0, 0
%	
%	qr = sae_new_residue(curve, true)
%	qnr = sae_new_residue(curve, false)
%	
%	# Set up the password and a dummy
%	base = bytearray(len(pwd))
%	dummy = get_random(len(pwd))
%	
%	for counter in range(1,41):
%		# Constant memory access version of base = found ? dummy : password;
%		# were the value is copied into base
%		const_time_select(found, dummy, password, base)
%		seed = H(max(a, b), min(a, b), base, counter)
%		# KDF handles gracefully the case x_cand > curve.p
%		x_cand = KDF(seed, "SAE Hunting and Pecking", curve.p)
%		
%		# res = 1 or 0 depending whether x_cand is valid or not
%		res = is_quadradic_residue(curve, x_cand, qr, qnr)
%		const_time_select_bin(found, x, x_cand, x)
%		save = const_time_select(found, save, seed[-1] & 0x01)
%		
%		# found is 0 or 0xff here and res is 0 or 1. Bitwise OR of them
%		# (with res converted to 0/0xff) handles this in constant time.
%		found |= res * 0xff
%		
%	# save is used to chose the value of y
%	pwe = point_from_bin(curve, x, save)
%	return pwe
%\end{lstlisting}

\subsection{Recommendations for Mitigations} \label{subsec:countermeasures}
Following the disclosure of Dragonblood, several mitigations have been 
proposed~\cite{IEEE_P802.11-19/1173r0,draft_harkins_eap_pwd_prime_00} to replace the iterative
 hash-to-group function by a deterministic function. This countermeasure suits our requirements. 
 However, backward compatibility might be a requirement in industry. Hence, we suggest to use a branch-free 
 implementation of the loop in order to avoid any residual leakage.

\begin{center}
\noindent \includegraphics[width=\columnwidth]{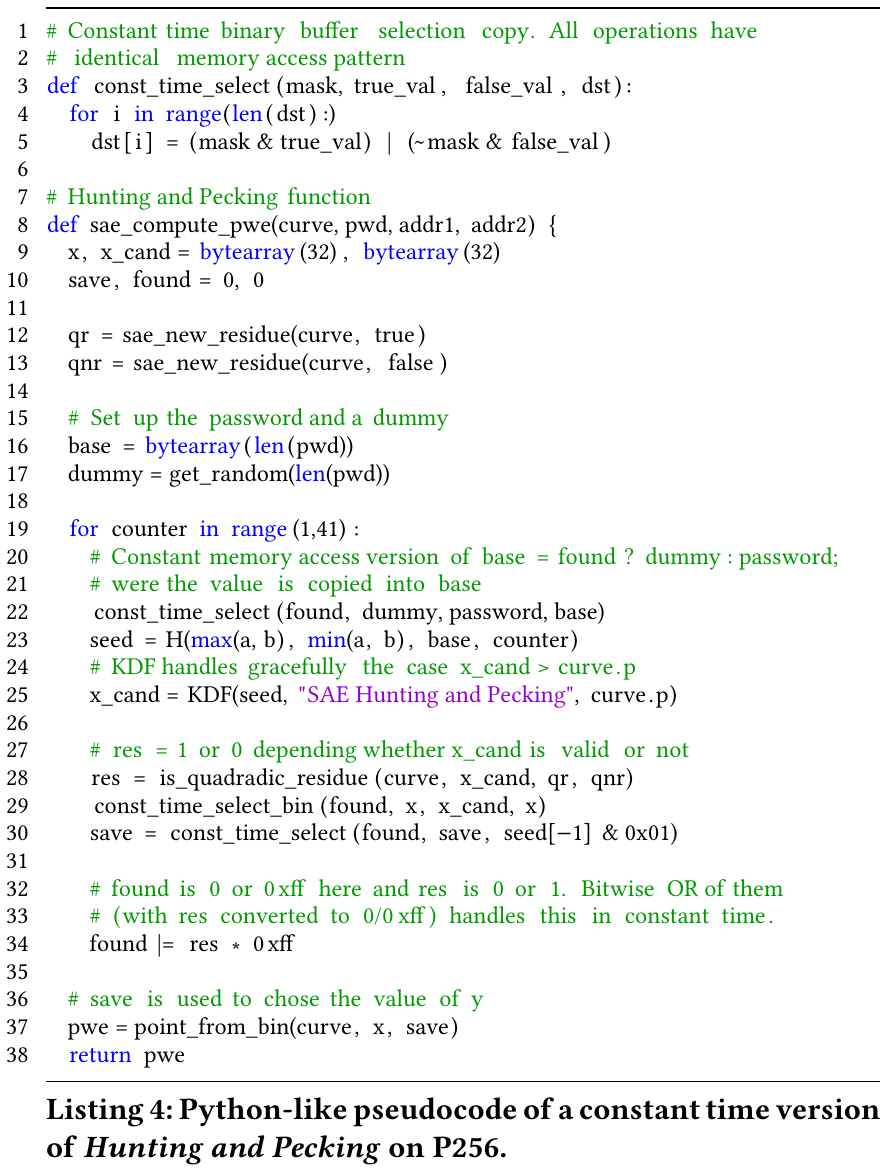}
\end{center}

We implemented such mitigations into iwd (see Listing~4), inspiring 
ourselves from hostapd patch\footnote{https://w1.fi/security/2019-1/}. We estimated the overhead 
induced by such countermeasure using the \texttt{rdtsc} assembly instruction, which offers very high 
precision. We made 10,000 measurements for both the mitigated derivation and the original one, 
while varying the password. We observed a negligible overhead ($1.4\cdot 10^{-9}\%$ on average).
The code complexity is barley affected by our changes. Considering the attack impact and the negligible downside of the patch, we strongly recommend  developers to include it in their products. Following our discoveries, both iwd and FreeRADIUS has smoothly integrated our patch in their code.

\subsection{Discussion} 
After the original Dragonblood publication, implementations received various patches, and dropped the support of some curves (mainly Brainpool curves). However, the main source of vulnerabilities, the hash-to-group function, is still unchanged, despite the standards update.

In spite of proper branch-free implementations being publicly available, with a negligible overhead, most implementations did not patch the secret-dependent control-flow of the password derivation. We believe the lack of patch is strongly related to the lack of Proof of Concept dedicated to specific implementations.  Dragonblood only describes the attack for hostapd which has been fixed.

We demonstrated that this vulnerability has more potential than the original one, allowing to recover more bits of information with fewer measurements. We provide a full Proof of Concept of our vulnerability on Intel's implementation, but we believe it can extend to others (see Appendix~\ref{app:freeradius}). Our approach illustrates the risk to users when cryptographic software developers dismiss a widely potential attack. This is unfortunately the prevailing approach for security vulnerabilities, but we show that for standards like WPA3, this approach is fraught with danger. Therefore, we hope that the Wi-Fi Alliance would drop their ad-hoc mitigations, for constant-time algorithms by design that do not rely on savvy developers to provide secure implementations. The history of PKCS\#1 v1.5 (with the Bleichenbacher attacks) shows that such a path is full of risks.  

%We intend to study other implementations, develop PoCs, and address the issues in a future work.

%!TEX root = article.tex

\begin{acks}
Daniel De Almeida Braga is funded by the Direction G\'en\'erale de l'Armement (P\^ole de Recherche CYBER). We would like to thank the anonymous paper and artifact reviewers for their time and constructive feedbacks.
\end{acks}

\bibliographystyle{ACM-Reference-Format}
\bibliography{article.bib}


\begin{thebibliography}{40}

%%% the Web address.

\ifx \showCODEN    \undefined \def \showCODEN     #1{\unskip}     \fi
\ifx \showDOI      \undefined \def \showDOI       #1{#1}\fi
\ifx \showISBNx    \undefined \def \showISBNx     #1{\unskip}     \fi
\ifx \showISBNxiii \undefined \def \showISBNxiii  #1{\unskip}     \fi
\ifx \showISSN     \undefined \def \showISSN      #1{\unskip}     \fi
\ifx \showLCCN     \undefined \def \showLCCN      #1{\unskip}     \fi
\ifx \shownote     \undefined \def \shownote      #1{#1}          \fi
\ifx \showarticletitle \undefined \def \showarticletitle #1{#1}   \fi
\ifx \showURL      \undefined \def \showURL       {\relax}        \fi
% invisible to TeX
\providecommand\bibfield[2]{#2}
\providecommand\bibinfo[2]{#2}
\providecommand\natexlab[1]{#1}
\providecommand\showeprint[2][]{arXiv:#2}

\bibitem[\protect\citeauthoryear{??}{778}{2016}]%
        {7786995}
 \bibinfo{year}{2016}\natexlab{}.
\newblock \showarticletitle{{IEEE Standard for Information
  technology—Telecommunications and information exchange between systems
  Local and metropolitan area networks—Specific requirements - Part 11:
  Wireless LAN Medium Access Control (MAC) and Physical Layer (PHY)
  Specifications}}.
\newblock \bibinfo{journal}{\emph{IEEE Std 802.11-2016 (Revision of IEEE Std
  802.11-2012)}} (\bibinfo{year}{2016}), \bibinfo{pages}{1--3534}.
\newblock


\bibitem[\protect\citeauthoryear{??}{rad}{2019}]%
        {radiator_security_advisory}
 \bibinfo{year}{2019}\natexlab{}.
\newblock \bibinfo{title}{{Two vulnerabilities in Radiator: EAP-pwd
  authentication bypass and DoS with certain TLS configurations}}.
\newblock
\newblock
\urldef\tempurl%
\url{https://open.com.au/OSC-SEC-2019-01.html}
\showURL{%
\tempurl}
\newblock
\shownote{Accessed: 2020-09-03.}


\bibitem[\protect\citeauthoryear{Aldaya, Garc{\'{\i}}a, Tapia, and
  Brumley}{Aldaya et~al\mbox{.}}{2019}]%
        {DBLP:journals/tches/AldayaGTB19}
\bibfield{author}{\bibinfo{person}{Alejandro~Cabrera Aldaya},
  \bibinfo{person}{Cesar~Pereida Garc{\'{\i}}a}, \bibinfo{person}{Luis
  Manuel~Alvarez Tapia}, {and} \bibinfo{person}{Billy~Bob Brumley}.}
  \bibinfo{year}{2019}\natexlab{}.
\newblock \showarticletitle{Cache-Timing Attacks on {RSA} Key Generation}.
\newblock \bibinfo{journal}{\emph{{IACR} Trans. Cryptogr. Hardw. Embed. Syst.}}
  \bibinfo{volume}{2019}, \bibinfo{number}{4} (\bibinfo{year}{2019}),
  \bibinfo{pages}{213--242}.
\newblock


\bibitem[\protect\citeauthoryear{Allan, Brumley, Falkner, van~de Pol, and
  Yarom}{Allan et~al\mbox{.}}{2016}]%
        {DBLP:conf/acsac/AllanBFPY16}
\bibfield{author}{\bibinfo{person}{Thomas Allan}, \bibinfo{person}{Billy~Bob
  Brumley}, \bibinfo{person}{Katrina~E. Falkner}, \bibinfo{person}{Joop van~de
  Pol}, {and} \bibinfo{person}{Yuval Yarom}.} \bibinfo{year}{2016}\natexlab{}.
\newblock \showarticletitle{Amplifying side channels through performance
  degradation}. In \bibinfo{booktitle}{\emph{{ACSAC}}}.
  \bibinfo{publisher}{{ACM}}, \bibinfo{pages}{422--435}.
\newblock


\bibitem[\protect\citeauthoryear{Alliance}{Alliance}{2019}]%
        {wpa3_security_considerations}
\bibfield{author}{\bibinfo{person}{Wi-Fi Alliance}.}
  \bibinfo{year}{2019}\natexlab{}.
\newblock \bibinfo{title}{{WPA3 Security Considerations}}.
\newblock
\newblock


\bibitem[\protect\citeauthoryear{Aranha, Novaes, Takahashi, Tibouchi, and
  Yarom}{Aranha et~al\mbox{.}}{2020}]%
        {DBLP:journals/iacr/AranhaNTTY20}
\bibfield{author}{\bibinfo{person}{Diego~F. Aranha},
  \bibinfo{person}{Felipe~Rodrigues Novaes}, \bibinfo{person}{Akira Takahashi},
  \bibinfo{person}{Mehdi Tibouchi}, {and} \bibinfo{person}{Yuval Yarom}.}
  \bibinfo{year}{2020}\natexlab{}.
\newblock \showarticletitle{LadderLeak: Breaking {ECDSA} With Less Than One Bit
  Of Nonce Leakage}.
\newblock \bibinfo{journal}{\emph{{IACR} Cryptol. ePrint Arch.}}
  \bibinfo{volume}{2020} (\bibinfo{year}{2020}), \bibinfo{pages}{615}.
\newblock


\bibitem[\protect\citeauthoryear{Bellardo and Savage}{Bellardo and
  Savage}{2003}]%
        {DBLP:conf/uss/BellardoS03}
\bibfield{author}{\bibinfo{person}{John Bellardo} {and} \bibinfo{person}{Stefan
  Savage}.} \bibinfo{year}{2003}\natexlab{}.
\newblock \showarticletitle{802.11 Denial-of-Service Attacks: Real
  Vulnerabilities and Practical Solutions}. In
  \bibinfo{booktitle}{\emph{{USENIX} Security Symposium}}.
  \bibinfo{publisher}{{USENIX} Association}.
\newblock


\bibitem[\protect\citeauthoryear{Benger, van~de Pol, Smart, and Yarom}{Benger
  et~al\mbox{.}}{2014}]%
        {DBLP:conf/ches/BengerPSY14}
\bibfield{author}{\bibinfo{person}{Naomi Benger}, \bibinfo{person}{Joop van~de
  Pol}, \bibinfo{person}{Nigel~P. Smart}, {and} \bibinfo{person}{Yuval Yarom}.}
  \bibinfo{year}{2014}\natexlab{}.
\newblock \showarticletitle{"Ooh Aah... Just a Little Bit" : {A} Small Amount
  of Side Channel Can Go a Long Way}. In \bibinfo{booktitle}{\emph{{CHES}}}
  \emph{(\bibinfo{series}{Lecture Notes in Computer Science})},
  Vol.~\bibinfo{volume}{8731}. \bibinfo{publisher}{Springer},
  \bibinfo{pages}{75--92}.
\newblock


\bibitem[\protect\citeauthoryear{Bruinderink, H{\"{u}}lsing, Lange, and
  Yarom}{Bruinderink et~al\mbox{.}}{2016}]%
        {DBLP:conf/ches/BruinderinkHLY16}
\bibfield{author}{\bibinfo{person}{Leon~Groot Bruinderink},
  \bibinfo{person}{Andreas H{\"{u}}lsing}, \bibinfo{person}{Tanja Lange}, {and}
  \bibinfo{person}{Yuval Yarom}.} \bibinfo{year}{2016}\natexlab{}.
\newblock \showarticletitle{Flush, Gauss, and Reload - {A} Cache Attack on the
  {BLISS} Lattice-Based Signature Scheme}. In
  \bibinfo{booktitle}{\emph{{CHES}}} \emph{(\bibinfo{series}{Lecture Notes in
  Computer Science})}, Vol.~\bibinfo{volume}{9813}.
  \bibinfo{publisher}{Springer}, \bibinfo{pages}{323--345}.
\newblock


\bibitem[\protect\citeauthoryear{Clarke and Hao}{Clarke and Hao}{2014}]%
        {DBLP:journals/iet-ifs/ClarkeH14}
\bibfield{author}{\bibinfo{person}{Dylan Clarke} {and} \bibinfo{person}{Feng
  Hao}.} \bibinfo{year}{2014}\natexlab{}.
\newblock \showarticletitle{Cryptanalysis of the dragonfly key exchange
  protocol}.
\newblock \bibinfo{journal}{\emph{{IET} Information Security}}
  \bibinfo{volume}{8}, \bibinfo{number}{6} (\bibinfo{year}{2014}),
  \bibinfo{pages}{283--289}.
\newblock


\bibitem[\protect\citeauthoryear{Cohney, Kwong, Paz, Genkin, Heninger, Ronen,
  and Yarom}{Cohney et~al\mbox{.}}{2020}]%
        {DBLP:conf/sp/CohneyKPGHRY20}
\bibfield{author}{\bibinfo{person}{Shaanan Cohney}, \bibinfo{person}{Andrew
  Kwong}, \bibinfo{person}{Shahar Paz}, \bibinfo{person}{Daniel Genkin},
  \bibinfo{person}{Nadia Heninger}, \bibinfo{person}{Eyal Ronen}, {and}
  \bibinfo{person}{Yuval Yarom}.} \bibinfo{year}{2020}\natexlab{}.
\newblock \showarticletitle{Pseudorandom Black Swans: Cache Attacks on
  CTR{\_}DRBG}. In \bibinfo{booktitle}{\emph{{IEEE} Symposium on Security and
  Privacy}}. \bibinfo{publisher}{{IEEE}}, \bibinfo{pages}{1241--1258}.
\newblock


\bibitem[\protect\citeauthoryear{Fluhrer}{Fluhrer}{2014}]%
        {IETF:mail_archive/cfrg/WXyM6pHDjGRZXZzSc_HlERnp0Iw}
\bibfield{author}{\bibinfo{person}{Scott Fluhrer}.}
  \bibinfo{year}{2014}\natexlab{}.
\newblock \bibinfo{title}{{Re: [CFRG] Requesting removal of CFRG co-chair}}.
\newblock
\newblock
\urldef\tempurl%
\url{https://mailarchive.ietf.org/arch/msg/cfrg/WXyM6pHDjGRZXZzSc_HlERnp0Iw/}
\showURL{%
\tempurl}


\bibitem[\protect\citeauthoryear{Fluhrer}{Fluhrer}{2018}]%
        {IETF:mail_archive/cfrg/mGnSNL8QW_fuCTwcyvh8lY9Z5G0}
\bibfield{author}{\bibinfo{person}{Scott Fluhrer}.}
  \bibinfo{year}{2018}\natexlab{}.
\newblock \bibinfo{title}{{Re: [Cfrg] I-D for password-authenticated EAP
  method}}.
\newblock
\newblock
\urldef\tempurl%
\url{https://mailarchive.ietf.org/arch/msg/cfrg/mGnSNL8QW_fuCTwcyvh8lY9Z5G0/}
\showURL{%
\tempurl}


\bibitem[\protect\citeauthoryear{Genkin, Valenta, and Yarom}{Genkin
  et~al\mbox{.}}{2017}]%
        {DBLP:conf/ccs/GenkinVY17}
\bibfield{author}{\bibinfo{person}{Daniel Genkin}, \bibinfo{person}{Luke
  Valenta}, {and} \bibinfo{person}{Yuval Yarom}.}
  \bibinfo{year}{2017}\natexlab{}.
\newblock \showarticletitle{May the Fourth Be With You: {A} Microarchitectural
  Side Channel Attack on Several Real-World Applications of Curve25519}. In
  \bibinfo{booktitle}{\emph{{ACM} Conference on Computer and Communications
  Security}}. \bibinfo{publisher}{{ACM}}, \bibinfo{pages}{845--858}.
\newblock


\bibitem[\protect\citeauthoryear{Harkins}{Harkins}{2014}]%
        {IEEE_P802.11-14/0640r1}
\bibfield{author}{\bibinfo{person}{Dan Harkins}.}
  \bibinfo{year}{2014}\natexlab{}.
\newblock \bibinfo{title}{{Addressing A Side-Channel Attack on SAE}}.
\newblock
\newblock
\urldef\tempurl%
\url{https://mentor.ieee.org/802.11/dcn/14/11-14-0640-01-000m-side-channel-attack.docx}
\showURL{%
\tempurl}


\bibitem[\protect\citeauthoryear{Harkins}{Harkins}{2015}]%
        {rfc7664}
\bibfield{author}{\bibinfo{person}{Dan Harkins}.}
  \bibinfo{year}{2015}\natexlab{}.
\newblock \bibinfo{title}{{Dragonfly Key Exchange}}.
\newblock \bibinfo{howpublished}{RFC 7664}.
\newblock
\urldef\tempurl%
\url{https://doi.org/10.17487/RFC7664}
\showDOI{\tempurl}


\bibitem[\protect\citeauthoryear{Harkins}{Harkins}{2019a}]%
        {IEEE_P802.11-19/1173r0}
\bibfield{author}{\bibinfo{person}{Dan Harkins}.}
  \bibinfo{year}{2019}\natexlab{a}.
\newblock \bibinfo{title}{{Finding PWE in Constant Time}}.
\newblock
\newblock
\urldef\tempurl%
\url{https://mentor.ieee.org/802.11/dcn/19/11-19-1173-08-000m-pwe-in-constant-time.docx}
\showURL{%
\tempurl}


\bibitem[\protect\citeauthoryear{Harkins}{Harkins}{2019b}]%
        {draft_harkins_eap_pwd_prime_00}
\bibfield{author}{\bibinfo{person}{D. Harkins}.}
  \bibinfo{year}{2019}\natexlab{b}.
\newblock \bibinfo{title}{Improved Extensible Authentication Protocol Using
  Only a Password draft-harkins-eap-pwd-prime-00}.
\newblock
\newblock
\urldef\tempurl%
\url{https://tools.ietf.org/html/draft-harkins-eap-pwd-prime-00}
\showURL{%
\tempurl}


\bibitem[\protect\citeauthoryear{Harkins}{Harkins}{2019c}]%
        {rfc8492}
\bibfield{author}{\bibinfo{person}{Dan Harkins}.}
  \bibinfo{year}{2019}\natexlab{c}.
\newblock \bibinfo{title}{{Secure Password Ciphersuites for Transport Layer
  Security (TLS)}}.
\newblock \bibinfo{howpublished}{RFC 8492}.
\newblock
\urldef\tempurl%
\url{https://doi.org/10.17487/RFC8492}
\showDOI{\tempurl}


\bibitem[\protect\citeauthoryear{Icart}{Icart}{2009}]%
        {DBLP:conf/crypto/Icart09}
\bibfield{author}{\bibinfo{person}{Thomas Icart}.}
  \bibinfo{year}{2009}\natexlab{}.
\newblock \showarticletitle{How to Hash into Elliptic Curves}. In
  \bibinfo{booktitle}{\emph{{CRYPTO}}} \emph{(\bibinfo{series}{Lecture Notes in
  Computer Science})}, Vol.~\bibinfo{volume}{5677}.
  \bibinfo{publisher}{Springer}, \bibinfo{pages}{303--316}.
\newblock


\bibitem[\protect\citeauthoryear{Igoe}{Igoe}{2012a}]%
        {IETF:mail_archive/cfrg/_BZEwEBBWhOPXn0Zw-cd3eSV6pY}
\bibfield{author}{\bibinfo{person}{Kevin~M. Igoe}.}
  \bibinfo{year}{2012}\natexlab{a}.
\newblock \bibinfo{title}{{[Cfrg] Status of DragonFly}}.
\newblock
\newblock
\urldef\tempurl%
\url{https://mailarchive.ietf.org/arch/msg/cfrg/_BZEwEBBWhOPXn0Zw-cd3eSV6pY/}
\showURL{%
\tempurl}


\bibitem[\protect\citeauthoryear{Igoe}{Igoe}{2012b}]%
        {IETF:mail_archive/cfrg/LsFX5Qqw53dTUmSsUOooLca5FHg}
\bibfield{author}{\bibinfo{person}{Kevin~M. Igoe}.}
  \bibinfo{year}{2012}\natexlab{b}.
\newblock \bibinfo{title}{{Re: [Cfrg] Status of DragonFly}}.
\newblock
\newblock
\urldef\tempurl%
\url{https://mailarchive.ietf.org/arch/msg/cfrg/LsFX5Qqw53dTUmSsUOooLca5FHg/}
\showURL{%
\tempurl}


\bibitem[\protect\citeauthoryear{{Intel Corporation}}{{Intel
  Corporation}}{2016}]%
        {IntelArchitectureOptimizationManual}
\bibfield{author}{\bibinfo{person}{{Intel Corporation}}.}
  \bibinfo{year}{2016}\natexlab{}.
\newblock \bibinfo{title}{{Intel® 64 and IA-32 Architectures Optimization
  Reference Manual}}.
\newblock
\newblock


\bibitem[\protect\citeauthoryear{Kügler}{Kügler}{2010}]%
        {IETF:mail_archive/ipsec/NEicYFDYJYcQuNdknY0etLyfITA}
\bibfield{author}{\bibinfo{person}{Dennis Kügler}.}
  \bibinfo{year}{2010}\natexlab{}.
\newblock \bibinfo{title}{{Re: [IPsec] PAKE selection: SPSK}}.
\newblock
\newblock
\urldef\tempurl%
\url{https://mailarchive.ietf.org/arch/msg/ipsec/NEicYFDYJYcQuNdknY0etLyfITA/}
\showURL{%
\tempurl}


\bibitem[\protect\citeauthoryear{Nik}{Nik}{2009}]%
        {rockyou}
\bibfield{author}{\bibinfo{person}{Cubrilovic Nik}.}
  \bibinfo{year}{2009}\natexlab{}.
\newblock \bibinfo{title}{{RockYou Hack: From Bad To Worse}}.
\newblock
\newblock
\urldef\tempurl%
\url{https://techcrunch.com/2009/12/14/rockyou-hack-security-myspace-facebook-passwords/}
\showURL{%
\tempurl}


\bibitem[\protect\citeauthoryear{{Nikolai Tschacher}}{{Nikolai
  Tschacher}}{2019}]%
        {Tschacher:2019}
\bibfield{author}{\bibinfo{person}{{Nikolai Tschacher}}.}
  \bibinfo{year}{2019}\natexlab{}.
\newblock \emph{\bibinfo{title}{{Model Based fuzzing of the WPA3 Dragonfly
  Handshake}}}.
\newblock \bibinfo{thesistype}{Master's\ thesis}. \bibinfo{school}{{Institute
  for Computer Science, Humboldt University}}, \bibinfo{address}{{Berlin,
  Germany}}.
\newblock


\bibitem[\protect\citeauthoryear{{NVlabs}}{{NVlabs}}{2016}]%
        {XMPlib}
\bibfield{author}{\bibinfo{person}{{NVlabs}}.} \bibinfo{year}{2016}\natexlab{}.
\newblock \bibinfo{title}{{XMP - CUDA accelerated(X) Multi-Precision library}}.
\newblock
\newblock
\urldef\tempurl%
\url{https://github.com/NVlabs/xmp}
\showURL{%
\tempurl}


\bibitem[\protect\citeauthoryear{Oren, Kemerlis, Sethumadhavan, and
  Keromytis}{Oren et~al\mbox{.}}{2015}]%
        {DBLP:conf/ccs/OrenKSK15}
\bibfield{author}{\bibinfo{person}{Yossef Oren}, \bibinfo{person}{Vasileios~P.
  Kemerlis}, \bibinfo{person}{Simha Sethumadhavan}, {and}
  \bibinfo{person}{Angelos~D. Keromytis}.} \bibinfo{year}{2015}\natexlab{}.
\newblock \showarticletitle{The Spy in the Sandbox: Practical Cache Attacks in
  JavaScript and their Implications}. In \bibinfo{booktitle}{\emph{{ACM}
  Conference on Computer and Communications Security}}.
  \bibinfo{publisher}{{ACM}}, \bibinfo{pages}{1406--1418}.
\newblock


\bibitem[\protect\citeauthoryear{Perrin}{Perrin}{2013}]%
        {IETF:mail_archive/tls/A_SfHI4BsdAi4miklBs3TvUbu-Y}
\bibfield{author}{\bibinfo{person}{Trevor Perrin}.}
  \bibinfo{year}{2013}\natexlab{}.
\newblock \bibinfo{title}{{[TLS] Review of Dragonfly PAKE}}.
\newblock
\newblock
\urldef\tempurl%
\url{https://mailarchive.ietf.org/arch/msg/tls/A_SfHI4BsdAi4miklBs3TvUbu-Y/}
\showURL{%
\tempurl}


\bibitem[\protect\citeauthoryear{Pessl, Bruinderink, and Yarom}{Pessl
  et~al\mbox{.}}{2017}]%
        {DBLP:conf/ccs/PesslBY17}
\bibfield{author}{\bibinfo{person}{Peter Pessl}, \bibinfo{person}{Leon~Groot
  Bruinderink}, {and} \bibinfo{person}{Yuval Yarom}.}
  \bibinfo{year}{2017}\natexlab{}.
\newblock \showarticletitle{To {BLISS-B} or not to be: Attacking strongSwan's
  Implementation of Post-Quantum Signatures}. In
  \bibinfo{booktitle}{\emph{{ACM} Conference on Computer and Communications
  Security}}. \bibinfo{publisher}{{ACM}}, \bibinfo{pages}{1843--1855}.
\newblock


\bibitem[\protect\citeauthoryear{Security}{Security}{[n.d.]}]%
        {crackstation:HumanPassword}
\bibfield{author}{\bibinfo{person}{Defuse Security}.}
  \bibinfo{year}{[n.d.]}\natexlab{}.
\newblock \bibinfo{title}{{CrackStation's Password Cracking Dictionary (Human
  Passwords Only}}.
\newblock
\newblock
\urldef\tempurl%
\url{https://crackstation.net/crackstation-wordlist-password-cracking-dictionary.htm}
\showURL{%
\tempurl}


\bibitem[\protect\citeauthoryear{van~de Pol, Smart, and Yarom}{van~de Pol
  et~al\mbox{.}}{2015}]%
        {DBLP:conf/ctrsa/PolSY15}
\bibfield{author}{\bibinfo{person}{Joop van~de Pol}, \bibinfo{person}{Nigel~P.
  Smart}, {and} \bibinfo{person}{Yuval Yarom}.}
  \bibinfo{year}{2015}\natexlab{}.
\newblock \showarticletitle{Just a Little Bit More}. In
  \bibinfo{booktitle}{\emph{{CT-RSA}}} \emph{(\bibinfo{series}{Lecture Notes in
  Computer Science})}, Vol.~\bibinfo{volume}{9048}.
  \bibinfo{publisher}{Springer}, \bibinfo{pages}{3--21}.
\newblock


\bibitem[\protect\citeauthoryear{Vanhoef and Piessens}{Vanhoef and
  Piessens}{2014}]%
        {DBLP:conf/acsac/VanhoefP14}
\bibfield{author}{\bibinfo{person}{Mathy Vanhoef} {and} \bibinfo{person}{Frank
  Piessens}.} \bibinfo{year}{2014}\natexlab{}.
\newblock \showarticletitle{Advanced Wi-Fi attacks using commodity hardware}.
  In \bibinfo{booktitle}{\emph{{ACSAC}}}. \bibinfo{publisher}{{ACM}},
  \bibinfo{pages}{256--265}.
\newblock


\bibitem[\protect\citeauthoryear{Vanhoef and Piessens}{Vanhoef and
  Piessens}{2017}]%
        {DBLP:conf/ccs/VanhoefP17}
\bibfield{author}{\bibinfo{person}{Mathy Vanhoef} {and} \bibinfo{person}{Frank
  Piessens}.} \bibinfo{year}{2017}\natexlab{}.
\newblock \showarticletitle{Key Reinstallation Attacks: Forcing Nonce Reuse in
  {WPA2}}. In \bibinfo{booktitle}{\emph{{ACM} Conference on Computer and
  Communications Security}}. \bibinfo{publisher}{{ACM}},
  \bibinfo{pages}{1313--1328}.
\newblock


\bibitem[\protect\citeauthoryear{Vanhoef and Ronen}{Vanhoef and Ronen}{2020}]%
        {DBLP:conf/sp/VanhoefR20}
\bibfield{author}{\bibinfo{person}{Mathy Vanhoef} {and} \bibinfo{person}{Eyal
  Ronen}.} \bibinfo{year}{2020}\natexlab{}.
\newblock \showarticletitle{Dragonblood: Analyzing the Dragonfly Handshake of
  {WPA3} and EAP-pwd}. In \bibinfo{booktitle}{\emph{{IEEE} Symposium on
  Security and Privacy}}. \bibinfo{publisher}{{IEEE}},
  \bibinfo{pages}{517--533}.
\newblock


\bibitem[\protect\citeauthoryear{Yarom}{Yarom}{2016}]%
        {Mastik}
\bibfield{author}{\bibinfo{person}{Yuval Yarom}.}
  \bibinfo{year}{2016}\natexlab{}.
\newblock \bibinfo{title}{{Mastik: A Micro-Architectural Side-Channel
  Toolkit}}.
\newblock
\newblock
\urldef\tempurl%
\url{https://cs.adelaide.edu.au/~yval/Mastik/}
\showURL{%
\tempurl}


\bibitem[\protect\citeauthoryear{Yarom and Benger}{Yarom and Benger}{2014}]%
        {DBLP:journals/iacr/YaromB14}
\bibfield{author}{\bibinfo{person}{Yuval Yarom} {and} \bibinfo{person}{Naomi
  Benger}.} \bibinfo{year}{2014}\natexlab{}.
\newblock \showarticletitle{Recovering OpenSSL {ECDSA} Nonces Using the
  {FLUSH+RELOAD} Cache Side-channel Attack}.
\newblock \bibinfo{journal}{\emph{{IACR} Cryptol. ePrint Arch.}}
  \bibinfo{volume}{2014} (\bibinfo{year}{2014}), \bibinfo{pages}{140}.
\newblock


\bibitem[\protect\citeauthoryear{Yarom and Falkner}{Yarom and Falkner}{2014}]%
        {DBLP:conf/uss/YaromF14}
\bibfield{author}{\bibinfo{person}{Yuval Yarom} {and} \bibinfo{person}{Katrina
  Falkner}.} \bibinfo{year}{2014}\natexlab{}.
\newblock \showarticletitle{{FLUSH+RELOAD:} {A} High Resolution, Low Noise,
  {L3} Cache Side-Channel Attack}. In \bibinfo{booktitle}{\emph{{USENIX}
  Security Symposium}}. \bibinfo{publisher}{{USENIX} Association},
  \bibinfo{pages}{719--732}.
\newblock


\bibitem[\protect\citeauthoryear{Yarom, Genkin, and Heninger}{Yarom
  et~al\mbox{.}}{2016}]%
        {DBLP:conf/ches/YaromGH16}
\bibfield{author}{\bibinfo{person}{Yuval Yarom}, \bibinfo{person}{Daniel
  Genkin}, {and} \bibinfo{person}{Nadia Heninger}.}
  \bibinfo{year}{2016}\natexlab{}.
\newblock \showarticletitle{CacheBleed: {A} Timing Attack on OpenSSL Constant
  Time {RSA}}. In \bibinfo{booktitle}{\emph{{CHES}}}
  \emph{(\bibinfo{series}{Lecture Notes in Computer Science})},
  Vol.~\bibinfo{volume}{9813}. \bibinfo{publisher}{Springer},
  \bibinfo{pages}{346--367}.
\newblock


\bibitem[\protect\citeauthoryear{Zorn and Harkins}{Zorn and Harkins}{2010}]%
        {rfc5931}
\bibfield{author}{\bibinfo{person}{Glen Zorn} {and} \bibinfo{person}{Dan
  Harkins}.} \bibinfo{year}{2010}\natexlab{}.
\newblock \bibinfo{title}{{Extensible Authentication Protocol (EAP)
  Authentication Using Only a Password}}.
\newblock \bibinfo{howpublished}{RFC 5931}.
\newblock
\urldef\tempurl%
\url{https://doi.org/10.17487/RFC5931}
\showDOI{\tempurl}


\end{thebibliography}

\appendix

\section{Password requiring more than 20 iterations on IWD}	\label{app:password_2X_iter}
%!TEX root = article.tex

Here is a sample of passwords requiring more than 20 iterations to be successfully derived into a point on P-256. 
MAC addresses are noted at the beginning of each list; the needed number of iterations is at the end of the line.

An extended list can be found in our gitlab
repository\footnote{https://gitlab.inria.fr/ddealmei/poc-iwd-acsac2020/-/blob/master/data/results/buggy\_passwords.txt}.

\begin{verbatim}
	## 992606B4AD9F FFF23027CB34 ##
	RAJARATNAM 21
	RA-KLEINENBERG 22
	ellochika 21
	VILIFYINGLY 24
	believeingod1 24
	BELLABOOBABE 25
	PRERRAFAELISTA 21
	DOGYLOVE1 21
	macarthurreviews 23
	AMERICANHOSPICE 21
	CHALLNENGE 22
	HAUNTEDEP 21
	Nibbler112 21
	0800581064 22
	SAKHLIKIS 21
	UPDMDFDr48 26
	kanakaman 30
	OXNWRABB35 23
	0874739218 23
	DEPEFCQ56 22
	taxidermically 21
	38concert 21
	NONPARISHIONER 22
	NOOMMAY7685 21
	gramocelj 21
	YUNKALLAH 21
	MILE-MICHEL-HYACINTHE 23
	STICKHANDLING 22
	faras-071196 21
	FARNHAM69 22
	10231976JR 21
	1102001625160 23
	wimjsbyk46 21
	veroleg351 27
	elasticized 21
	cutelildevilj87 21
	JLNRUJY98 25
	FENWICK-1994 21
	
	## CF116C758375 553ED5460AA7 ##
	th-commando-regiment 24
	thechildrensbank 21
	0143576155 22
	POWERTOHARM 22
	EMILYELAINE 22
	becks4svs 22
	wvdbincy98448342 25
	RCCB16023 22
	9117820114 24
	mbuyisa's 24
	islandinstlawrencewithducks 23
	volume-issue 21
	ASPINKK202 21
	jratlnve54 22
	s9040954i 21
	cerinek007 21
	JULIESULLIVAN 24
	DOXIE\_CHIC 21
	AujcYOLE24 22
	WALTHAMSTOWEAST 25
	tightrope-men 22
	FOODENGINEERINGMAG 26
	PROSTITITUES 22
	SHEAILY872264 22
	contest-win-weezers-boombox 21
	drkencarter 21
	UNIVERSALVEILING 21
	taka-taka 21
	0849852969 21
	otiwbawm61 21
	ouchana170672 23
	0860168289 22
	SIEDING63 21
	GORDON520P 26
	midmanhattan 21
	QgUPaKF67 21
	3THUGLOVE 21
	scarcetheband 30
	tegetiformans 21
	canadiancray 25
	egzistencija 21
	civilrecht 23
	BONGONITO 22
\end{verbatim}

\section{Sample of a Trace of IWD} \label{app:trace_sample}
%!TEX root = article.tex

Sample of a trace yielding four iterations. This has been acquired using the password \texttt{superpassword}, 
with MAC addresses \texttt{E2F754FE22D1} and \texttt{9203835A576B}.
Annotations have been added and are not part of the original trace.

\begin{verbatim}
# First five lines correspond to the qr and qnr generation
# They are ignored during parsing
l_getrandom 5435937 (90)
l_getrandom 5439791 (88)
l_getrandom 5443732 (96)
l_getrandom 5447611 (88)
l_getrandom 5455232 (88)
# Here the loop begins
kdf_sha256 5459308 (82)
kdf_sha256 3324 (86)
kdf_sha256 4091 (82)
kdf_sha256 3972 (84)
l_getrandom 108 (90)
# At the fourth iteration, we notice long-delayed call 
# to l_getrandom. It means we can stop there.
l_getrandom 3889 (88)
	
kdf_sha256 3981 (82)
kdf_sha256 4089 (84)
l_getrandom 106 (90)
kdf_sha256 3734 (86)
kdf_sha256 9058 (100)
kdf_sha256 417 (84)
l_getrandom 501 (90)
kdf_sha256 5691 (84)
l_getrandom 129 (94)
kdf_sha256 3795 (88)
# Other long-delayed calls can be observed, hence 
# the need to acquire multiple samples
l_getrandom 4320 (96)
kdf_sha256 4524 (86)
...
\end{verbatim}

\section{Attack on FreeRadius}\label{app:freeradius}
%!TEX root = article.tex

FreeRADIUS supports EAP-pwd, a variant of Dragonfly, as a non-default authentication method, encapsulated in the RADIUS protocol. Beside the patches to Dragonblood attacks, we show that EAP-pwd is still vulnerable to timing attacks (due to a variable number of iterations), and to the same cache attack we described in Section~\ref{sec:attack}. In this section, we studied the last version of FreeRADIUS (v3.0.21 at the time of writing).

\subsection{EAP-pwd vs SAE}

SAE and EAP-pwd being two variants of Dragonfly, they differ in a few points. Some of them are only instantiation details (values of some labels), while others have more impactful consequences on the workflow and the security of the protocol.

First, EAP-pwd standard does not mandate a constant number of iterations. Indeed, it exits the conversion loop as soon as the password is successfully converted. Since a constant number of iterations would not change the outcome of the conversion, some implementations (not FreeRADIUS) include this side-channel mitigation anyway.

Next, EAP-pwd does not benefit from the same symmetry as SAE: client and server are clearly defined. This distinction is highlighted by the fact that the server generates a random token for each new session. This token will be part of the information hashed at each iteration during the password conversion.
Hence, while a password is always derived into the same element in SAE (as long as the identities do not change), each EAP-pwd session ends up with a new group element, due to the randomness brought by the token.

\subsection{FreeRadius implementation}

The Dragonfly exchange implemented by FreeRadius follows EAP-pwd's specification~\cite{rfc5931}. All related functions are defined in the according module\footnote{\url{https://github.com/FreeRADIUS/freeradius-server/tree/v3.0.x/src/modules/rlm\_eap/types/rlm\_eap\_pwd}}. Namely, the \emph{Hunting and Pecking} is implemented in the function \texttt{compute\_password\_element}, as illustrated in Listing~5. We cut some parts of the code, and renamed variables for the sake of clarity.

\begin{center}
\noindent \includegraphics[width=\columnwidth]{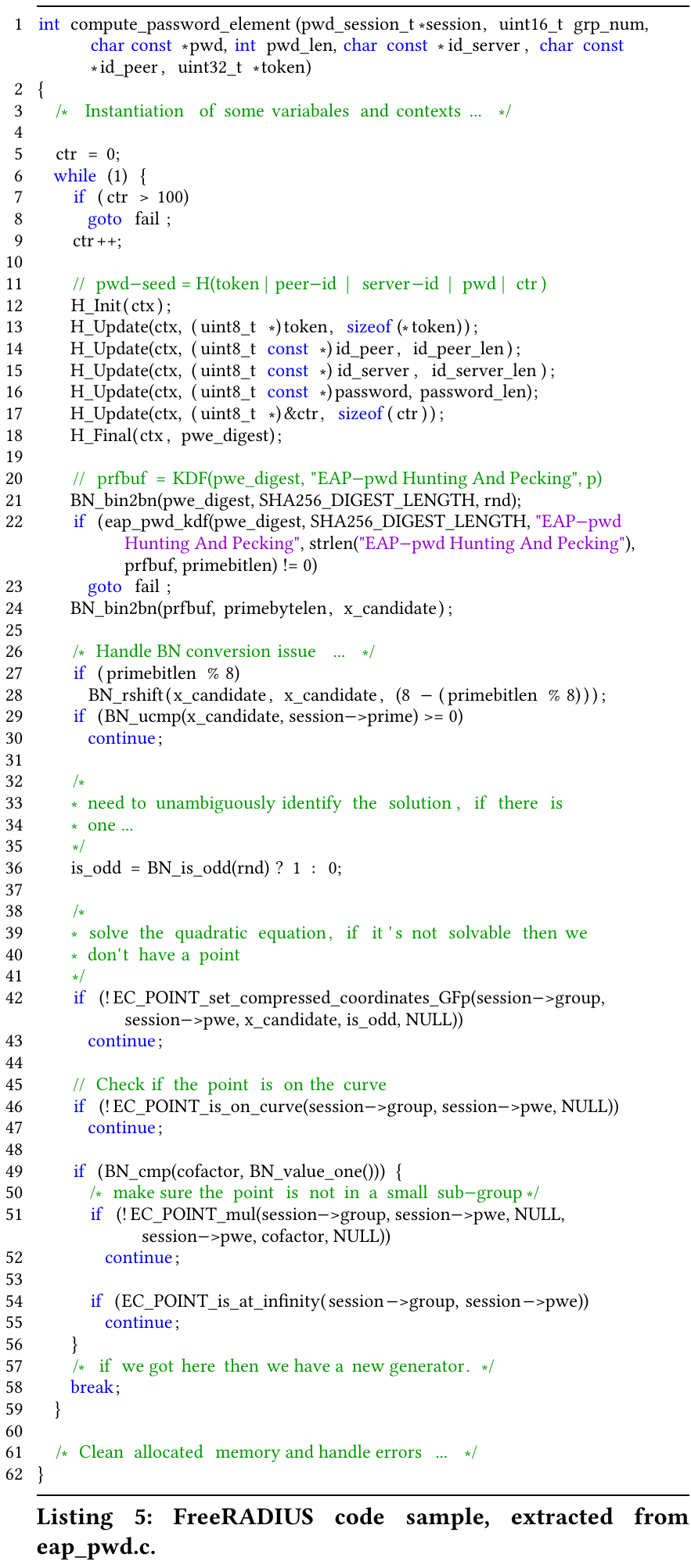}
\end{center}

This implementation heavily relies on OpenSSL\footnote{https://www.openssl.org/} to perform cryptographic operations, such as hashing, manipulating big integers and elliptic curve points. By default, the library is dynamically linked from the system-wide installation when building the project.

A quick look at the code in Listing~5 shows a few branches inside the loop. At line 30, the iteration will end if the output of the KDF is bigger than the prime. At line 43, if the candidate is not an x-coordinate of a point on the curve, the rest of the loop is skipped. The same phenomenon occurs at line 47 and 52. Finally, at line 58, the loop ends if a password have been found, making the total number of operation password-dependent.

Since the issue of having a password-dependent number of iteration (yielding a clear timing difference) has already been discussed in~\cite{DBLP:conf/sp/VanhoefR20}, we will focus on the cache attack allowing to guess the exact number of iterations needed to convert the password, even if the total number of iterations is fixed.

\subsection{Cache-Attack Against FreeRADIUS}

Using some minor adaptations, we applied our cache attack (described on iwd in  Section~\ref{sec:attack}) to guess the exact iteration in which the password is successfully derived. We stress that switching to a constant number of iterations, with a constant time (or masked) Legendre symbol computation, would mitigate the timing attack, but our cache attack would still be practical.

We perform this attack by only monitoring two memory lines, both in the OpenSSL cryptographic library. To do so, we use the calls to \texttt{H\_Update} 
(called line 13 to 17) as a synchronization clock. Since multiple calls to this function follow each other, we catch them with high probability.
Next, we use the call to \texttt{EC\_POINT\_is\_on\_curve} (line 46) as a success-specific code. More specifically, this function calls \texttt{set\_affine\_coordinates} from OpenSSL internals, which is also called if the original check (line 42) is successful success. Thus, some piece of code is called twice on success, and is never called on failure.

\subsection{Experimental results}

We implemented a full Proof of Concept of our attack, and made it publicly available\footnote{https://gitlab.inria.fr/msabt/attack-poc-freeradius} after the vulnerability has been patched. The experimental setup is the same as described in Section~\ref{ssec:setup}.

Due to the server-generated token, we only have a single measurement to guess how many iterations are needed to convert the password. We tested our attack on 80 different passwords, spying on 15 connections for each password, yielding a total of 1200 traces. With a single measurement, we successfully guessed the exact number of iterations for 93\% of the traces. We outline some consistency in the errors: most errors occurred because the spy process misses on call to the synchronization clock. 
Hence, we can achieve a better reliability by loosing some information: assuming that if we guess that the password needs $x$ iterations to be converted, then it may need $x$ or $x+1$ iterations, allowing us to reach 99\% accuracy.

Considering we achieve this accuracy with a single measurement, we are able to recover a password with fewer measurements than in previous attacks, even by softening our guess.

\end{document}